\documentclass[%
superscriptaddress,
 amsmath,amssymb,
 aps,
prfluids,
]{revtex4-2}

\usepackage{float}
\usepackage{booktabs,array}
\usepackage{graphicx}% Include figure files
\usepackage{dcolumn}% Align table columns on decimal point
\usepackage{bm}
\usepackage{xcolor}
\usepackage{soul}% bold math
\newcommand\Real{\mbox{Re}}
 
\newcommand\Rey{\mbox{\textit{Re}}}  % Reynolds number
\begin{document}

\preprint{APS/123-QED}

\title{Energy transfer mechanisms and resolvent analysis in the cylinder wake}% Force line breaks with \\
%\thanks{A footnote to the article title}%

\author{Bo Jin}
\email{bjin1@student.unimelb.edu.au}
% \altaffiliation[Also at ]{Department of Mechanical Engineering, University of Melbourne,VIC, 3010, Australia}%Lines break automatically or can be forced with \\
\author{Sean Symon}%
%\email{Second.Author@institution.edu}
\author{Simon J. Illingworth}%
%\email{Second.Author@institution.edu}
\affiliation{%
 Department of Mechanical Engineering \\University of Melbourne,VIC, 3010, Australia
}%

% \collaboration{MUSO Collaboration}%\noaffiliation

% \author{Charlie Author}
%  \homepage{http://www.Second.institution.edu/~Charlie.Author}
% \affiliation{
%  Second institution and/or address\\
%  This line break forced% with \\
% }%
% \affiliation{
%  Third institution, the second for Charlie Author
% }%
% \author{Delta Author}
% \affiliation{%
%  Authors' institution and/or address\\
%  This line break forced with \textbackslash\textbackslash
% }%

% \collaboration{CLEO Collaboration}%\noaffiliation

\date{\today}% It is always \today, today,
             %  but any date may be explicitly specified
\begin{abstract}
%discrepancies between the true flow field and that predicted by resolvent analysis.
\textcolor{black}{We investigate energy-transfer mechanisms for vortex shedding in the two-dimensional cylinder wake at a Reynolds number of $\Rey=100$. In particular, we focus on a comparison of the energy transfer in the true flow field to that predicted by resolvent analysis.} The energy balances achieved by the true cylinder flow are first characterized---both for the flow as a whole and for each of its most energetic harmonic frequencies. It is found that viscous dissipation balances production when each is considered over the entire flow field and therefore that linear mechanisms achieve an energy balance on their own, thus respecting the Reynolds--Orr equation. Suitable energy conservation laws reveal that while nonlinear mechanisms neither produce nor consume energy overall, they nevertheless account for an important transfer of energy to higher frequencies. The energy balance for DNS is compared to that predicted by resolvent analysis. Although a suitable energy balance is achieved for each harmonic, resolvent analysis does not correctly model nonlinear energy transfer between temporal frequencies \textcolor{black}{since it only models interactions between the mean flow and fluctuations. The impact of the neglected nonlinear energy transfer on the resolvent mode shapes is also made clear by analyzing the spatial distribution of the energy transfer mechanisms.} We then investigate the detailed roles of energy transfer from the viewpoint of nonlinear triadic interactions by considering a finite number of harmonic frequencies. It is shown that inter-fluctuation interactions play a critical role in the redistribution of energy to higher harmonic frequencies. We observe not only an energy cascade from low frequencies to high frequencies, but also a considerable inverse cascade from high frequencies to low frequencies. \textcolor{black}{The true energy pathways observed among harmonic modes provide additional constraints that could help to improve the modeling of nonlinear energy transfer in the cylinder flow.}
\end{abstract}

%\keywords{Suggested keywords}%Use showkeys class option if keyword
                              %display desired
\maketitle

%\tableofcontents
\section{Introduction}
Shear flows occur whenever a fluid flows past a solid object and are therefore commonplace in engineering and in nature.
Despite being governed by the Navier--Stokes equations---a set of nonlinear partial differential equations---certain important aspects of shear flows are well described by linear mechanisms \citep{Schmid01}. This is true not only for small perturbations away from some laminar base flow, but also for unsteady flows for which the fluctuations are not small. 
%\textcolor{black}{Therefore, quasi-linear theory has been employed to reduce the complexity of nonlinear fluid flows and approximate the dynamics of large scale structures \citep{marston2016generalized}.}
For unsteady flows, a linear operator can be formed about the time-averaged mean flow and the remaining nonlinear terms are then treated as a forcing to an otherwise linear system \citep{landahl1967wave,bark1975wave}.
In this way no linearization is performed.
Rather one characterizes the response of the linear operator to the remaining nonlinear terms.
In this context resolvent analysis---in which the linear operator is characterized in the frequency domain by forming its temporal frequency response (its resolvent)---has been used with particular success in recent years \citep{McKeon10}.
Despite its success, some of the predictions of resolvent analysis show important discrepancies with the true flow.
Of particular note for the present work is that \textit{i}) the predictions of resolvent analysis are often improved by including an eddy viscosity in the linear operator \citep{Hwang10, Mettot14, Illingworth18}; and that \textit{ii}) for some shear flows such as the cylinder wake \citep{Rosenberg19} and the flow past an airfoil \citep{Yeh19, Symon19}, the leading resolvent response mode is too energetic in the far wake.

This work considers resolvent analysis and energy transfer mechanisms for the two-dimensional flow past a cylinder at a Reynolds number of 100. Resolvent analysis of the mean flow reveals a dominant linear mechanism due to a marginally stable eigenvalue whose frequency matches the vortex shedding \cite{symon2018non,Barkley06}. Even though the flow contains harmonics of the vortex shedding frequency, the Reynolds stresses are primarily comprised of the fundamental harmonic interacting with its complex conjugate \cite{Sipp07}. As such, a self-consistent description of the flow can be obtained by considering a truncated set of equations between the mean flow and the fundamental harmonic \cite{mantivc2014self}. Nevertheless, linear analysis does not correctly model the streamwise decay of the vortex shedding and fails to predict the spatial structure of higher harmonics. \textcolor{black}{These discrepancies make resolvent modes an inefficient basis for practical applications such as reduced-order modeling and flow estimation.}
	
In this paper, we perform an harmonic decomposition of the cylinder wake and characterize linear and nonlinear energy transfer mechanisms. The true energy transfer from direct numerical simulation (DNS) is compared to that predicted by a resolvent analysis of each harmonic mode. By doing so we characterize---both for the true flow and for resolvent analysis---the energy balance between production, dissipation, and nonlinear transfer for each harmonic mode. We show that, although resolvent analysis achieves a suitable energy balance for any single harmonic mode, it does not achieve the correct energy balance across modes. In particular we will observe a significant nonlinear transfer of energy from the first harmonic mode to the second and third harmonic modes in DNS which is not captured by resolvent analysis. Additionally, we analyze all nonlinear exchanges of energy among the harmonics that play a role in sustaining the flow.

The rest of the paper is organized as follows. In Sec.~\ref{sec:problem formulation}, we briefly review resolvent analysis and describe the direct numerical simulation (DNS) of cylinder flow. In Sec.~\ref{sec:methods}, we derive the energy balances that must be satisfied for cylinder flow overall and for each harmonic mode. In Sec.~\ref{sec:result1}, we characterize, for the DNS and the leading resolvent mode, the energy transfer mechanisms for each harmonic and the flow overall. We also compare the spatial distribution of each term in the energy balance as computed from the DNS and the leading resolvent mode. We will see that the excess energy of the leading mode in the far wake \cite{Yeh19, Rosenberg19} can be attributed to resolvent analysis incorrectly predicting nonlinear transfer. In Sec.~\ref{sec:result2}, we investigate energy exchanges through triadic interactions among the various harmonics and discuss the implications of our results for quasi-linear modeling applications. Finally, we conclude in Sec.~\ref{sec:conclusion}.

\section{Problem formulation} \label{sec:problem formulation}
\textcolor{black}{The objective of this study is to investigate discrepancies between the true flow field and the quasi-linear model (i.e.~resolvent modes) from an energy transfer perspective. In particular, we focus on two aspects: i) recovering the true energy transfer pathways across frequencies and ii) quantifying the contribution of inter-fluctuation interactions to the energy balance. This provides a benchmark for evaluating the extent to which resolvent analysis can correctly model energy transfer mechanisms, which is essential for improving the modeling and simulation of nonlinear fluid flows. In Sec.~\ref{sec:II.A}, we describe the governing equations for incompressible fluid flows and the evolution of their fluctuations about the time-averaged mean. A brief introduction to linear input-output (resolvent) analysis is provided in Sec.~\ref{sec:II.B}. In Sec.~\ref{sec:II.C}, the results of the resolvent analysis are presented and compared to those obtained from direct numerical simulations (DNS).}

\subsection{Incompressible nonlinear flow}\label{sec:II.A}
The incompressible Navier-Stokes equations describe the conservation of mass and momentum of an incompressible fluid:
\begin{equation}\label{equ:nsequation}
    \begin{gathered}
        \partial_t\textbf{\textit{u}}=-\textbf{\textit{u}}\cdot\nabla\textbf{\textit{u}}-\nabla \textit{p}+\Rey^{-1} \nabla^2\textbf{\textit{u}},\\
        \nabla\cdot\textbf{\textit{u}}=0.
    \end{gathered}
\end{equation}
The equations have been nondimensionalized by the free-stream velocity $U_{\infty}$ and the cylinder diameter $D$.
The Reynolds number is therefore defined as $\Rey=U_{\infty} D/\nu$, where $\nu$ is the kinematic viscosity.
For a fully developed flow, an evolution equation can be derived for the fluctuations ($\textbf{\textit{u}}',\ \textit{p}'$) by forming a linear operator about the mean (time-averaged) flow ($\textbf{\textit{U}},\ \textit{P}$) and treating the remaining nonlinear terms as an exogenous forcing \citep{landahl1967wave,bark1975wave,McKeon10}:
\begin{equation}\label{equ:pertur}
  \begin{gathered}
    \partial_t\textbf{\textit{u}}'=L\textbf{\textit{u}}'-\nabla\textit{p}'+\textbf{\textit{f}}',\\
    \nabla\cdot\textbf{\textit{u}}'=0.
  \end{gathered}
\end{equation}
Thus the fluctuations evolve according to the linear operator $L = {-\textbf{\textit{U}}\cdot\nabla()} - {()\cdot\nabla\textbf{\textit{U}}} + {\Rey^{-1}\nabla^2()}$ and the forcing they receive from the remaining nonlinear terms $\textbf{\textit{f}}' = {-\textbf{\textit{u}}'\cdot\nabla\textbf{\textit{u}}'} + {\overline{\textbf{\textit{u}}'\cdot\nabla\textbf{\textit{u}}'}}$.
(Here $\overline{(\cdot)}$ denotes a time average.)
Note that the pressure term in Eq.~\eqref{equ:pertur} could be eliminated by projecting the velocity field onto the space of divergence-free functions, but we choose to retain pressure for the moment because its influence will be eliminated naturally when we consider the flow's energy balance in Sec.\,\ref{sec:methods}.

\subsection{Linear input/output (resolvent) analysis}\label{sec:II.B}
\textcolor{black}{The remaining nonlinear forcing poses considerable difficulties for linear modeling approaches. To address this problem, Eq.~\eqref{equ:pertur} can be reframed into a linear input-output form, where the nonlinear term $\textbf{\textit{f}}'$ is treated as an intrinsic forcing to the system and generates a velocity field through the resolvent operator that is linearised around the mean flow. The spectral properties of the resolvent operator reflect important information about linear amplification mechanisms, e.g.~the most dangerous disturbances and the velocity field to which they give rise.} 

Resolvent analysis proceeds by taking Laplace transforms of Eq.~\eqref{equ:pertur}, setting $s=j\omega$ and rearranging to arrive at the frequency response (or resolvent):
\begin{equation}\label{equ:linearoperator}
  \mathcal{H}(\omega)=\textcolor{black}{\textbf{R}^T}\left( j\omega\textbf{I}+\begin{bmatrix}
    -L&\nabla ()\\
    \nabla\cdot()&0
    \end{bmatrix}\right)^{-1}\textcolor{black}{\textbf{R}},
\end{equation}
so that, at frequency $\omega$,
\begin{equation}
  \hat{\textbf{\textit{u}}}(\omega) = \mathcal{H}(\omega) \hat{\textbf{\textit{f}}}(\omega).
\end{equation}
%$\hat{\textbf{\textit{u}}}=\mathcal{H}(j\omega)\hat{\textbf{\textit{f}}}$
\textcolor{black}{Here $\textbf{R}=[\textbf{I},\ \textbf{0}]^{T}$ is the prolongation matrix that consists of an identity block matrix for the velocity and a zero block matrix for the pressure. Therefore, $\textbf{R}$ maps a velocity vector $\hat{\textbf{\textit{u}}}$ to a velocity-zero-pressure vector $[\hat{\textbf{\textit{u}}},\ \textbf{0}]^{T}$ whereas $\textbf{R}^T$ maps a velocity-pressure vector $[\hat{\textbf{\textit{u}}},\ \hat{\textit{p}}]^{T}$ to a velocity vector $\hat{\textbf{\textit{u}}}$.}
Rather than consider the exact form of the nonlinear forcing $\hat{\textbf{\textit{f}}}$, resolvent analysis instead considers the optimal forcing that achieves the maximum energetic gain $\gamma^2(\omega)$ at each frequency. \textcolor{black}{In other words, we consider only the optimal resolvent forcing and response modes at each frequency, as an optimal rank-1 approximation of the system's frequency response.}
This analysis gives the most important information on the system's linear dynamics by characterizing its global frequency response to external forcing \citep{sipp2013characterization}.
The linear optimization to be performed is
\begin{equation}\label{equ:resolvent_opti}
  \gamma^2(\omega) 
  =
  \max_{\hat{\textbf{\textit{f}}}}\hspace{1mm}\dfrac{\langle\hat{\textbf{\textit{u}}}\hspace{0mm}^*,\hat{\textbf{\textit{u}}}\rangle}{\langle\hat{\textbf{\textit{f}}}\hspace{0mm}^*,\hat{\textbf{\textit{f}}}\rangle}
  =
  \max_{\hat{\textbf{\textit{f}}}}\hspace{1mm}\dfrac{\langle\hat{\textbf{\textit{f}}}\hspace{0mm}^*\mathcal{H}^*(\omega),\mathcal{H}(\omega)\hat{\textbf{\textit{f}}}\rangle}{\langle\hat{\textbf{\textit{f}}}\hspace{0mm}^*,\hat{\textbf{\textit{f}}}\rangle}=\sigma_1^2(\mathcal{H}(\omega)),
\end{equation}
where $\langle\cdot,\cdot\rangle$ denotes the inner product over the spatial domain $\Omega$.
This optimization problem can be solved by performing a singular value decomposition of the resolvent operator $\mathcal{H}(\omega)$: that is, the leading singular value squared, $\sigma_1^2(\mathcal{H}(\omega))$, corresponds to the maximum energy gain $\gamma^2(\omega)$ at frequency $\omega$, as expressed by Eq. \eqref{equ:resolvent_opti}. 

\subsection{DNS and linear prediction}\label{sec:II.C} %\label{sec:config}
\begin{figure}
    \centerline{\includegraphics[width=0.45\textwidth]{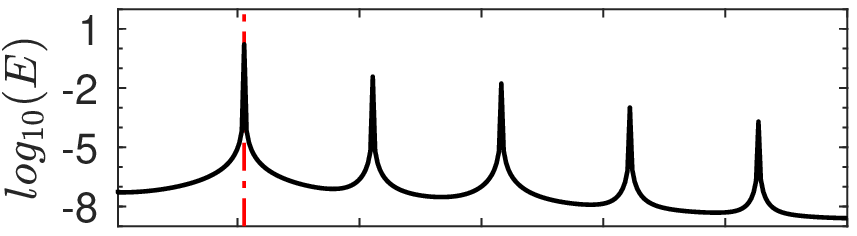}
        \llap{\parbox[b]{6.6in}{(a)\\\rule{0ex}{0.8in}}}
        \hspace{5mm}
    \includegraphics[width=0.43\textwidth,bb=0 296 645 288]{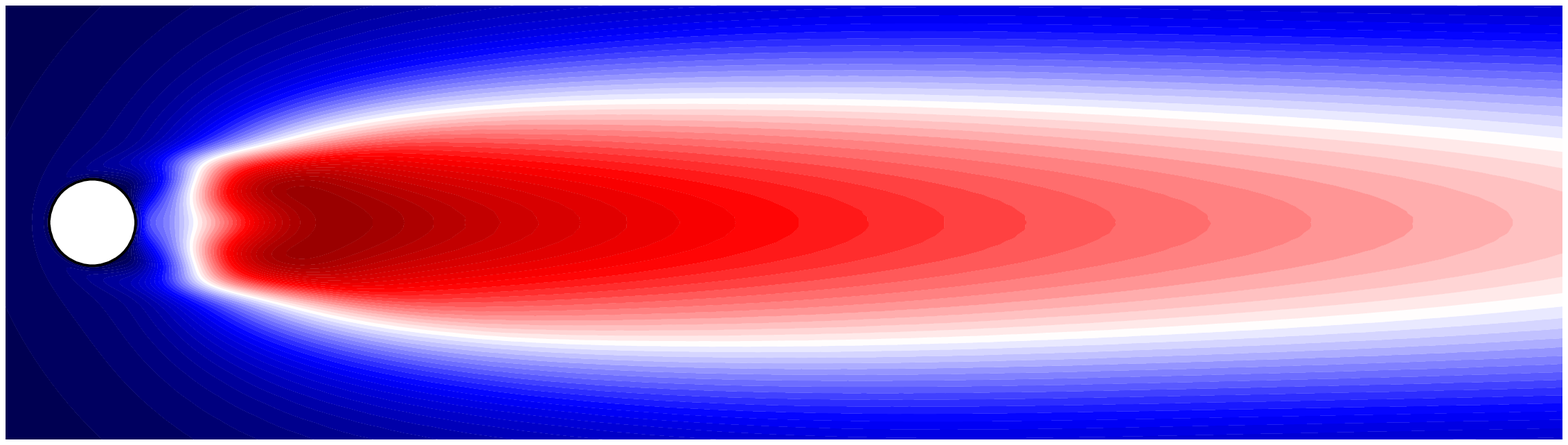}
		\llap{\parbox[b]{6.55in}{(c)\\\rule{0ex}{0.8in}}}
    }
    \vspace{1mm}
	\centerline{\includegraphics[width=0.45\textwidth]{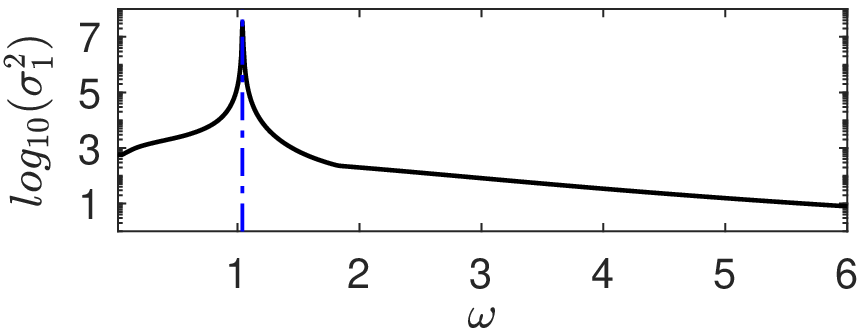}
		\llap{\parbox[b]{6.6in}{(b)\\\rule{0ex}{1.1in}}}
		\hspace{5mm}
	\includegraphics[width=0.43\textwidth,bb=0 234 645 288]{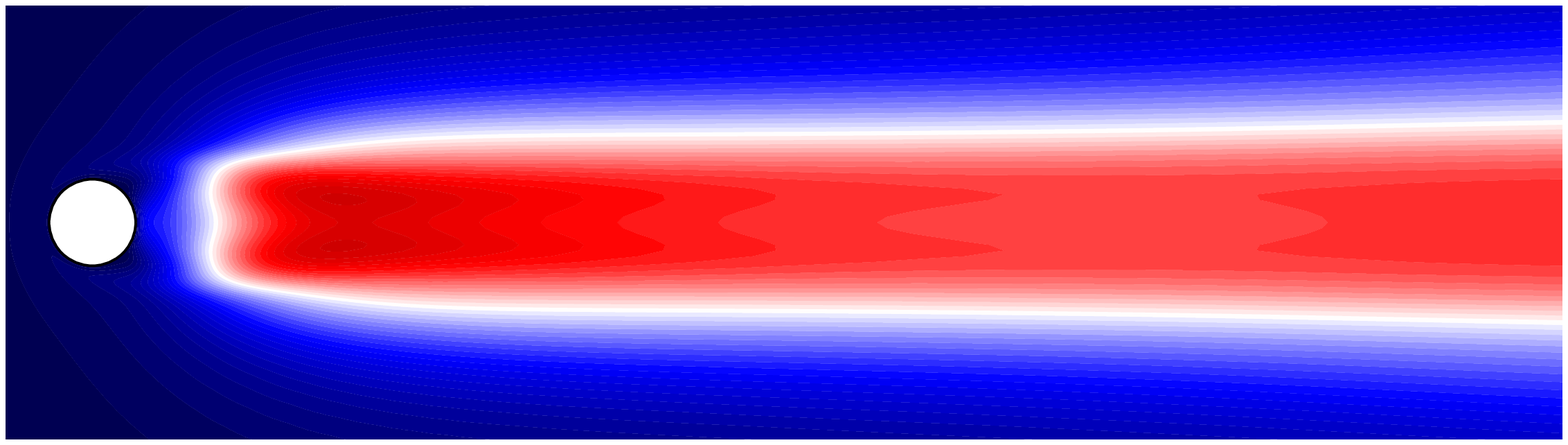}
		\llap{\parbox[b]{6.55in}{(d)\\\rule{0ex}{1.1in}}}
	}
	\caption{$(\text{a})$ Energy spectrum of perturbation velocity from DNS. $(\text{b})$ Leading energy gains $\sigma_1^2$ from resolvent analysis. Spatial distribution of the kinetic energy for $(\text{c})$ 1st harmonic mode from DNS and $(\text{d})$ optimal response at 1st harmonic frequency from resolvent analysis. Both contour plots share the same color scale.}
	\label{fig:dnsresolvent}
\end{figure}

\textcolor{black}{We now consider direct numerical simulations and linear analyses of the incompressible flow past a two-dimensional circular cylinder at a Reynolds number of $\Rey=100$, both of which have been performed using the FEniCS platform \citep{logg2012automated}.} The Navier-Stokes equations are discretized using Taylor-Hood finite elements over a structured mesh. We employ the same computational domain as that used in \citep{leontini2006wake} and the corresponding boundary conditions have been tested in \cite{jin_illingworth_sandberg_2020}.

We simulate the full nonlinear equations Eq.~\eqref{equ:nsequation} using a time step $\Delta t=0.01$.
A second-order implicit scheme is used for time discretization and the resulting nonlinear equations are solved directly using a Newton method.
The simulations give rise to vortex shedding at a Strouhal number of $\text{St}=0.1677$.
This is consistent with the results of \cite{jiang2017strouhal} and corresponds to a fundamental frequency of $\omega_1=1.054$.
The time-averaged mean flow is then obtained by time-averaging over 54 periods after the flow has settled to saturated vortex shedding, and the fluctuations are extracted for Fourier analysis.
The corresponding energy spectrum is shown in Fig.~\ref{fig:dnsresolvent}(\textit{a}), which displays sharp peaks at the fundamental frequency $\omega_1$ (marked by a red dashed line) and its higher harmonics $\omega_n=n\omega_1$. \textcolor{black}{Here, $n\in\mathbb{Z}^{+}$ is a positive integer.}
It is important to note that more than $99.9\%$ of the total fluctuation energy is concentrated in the first three harmonic frequencies $\omega_1$, $\omega_2$, and $\omega_3$.

We also characterize in Fig.~\ref{fig:dnsresolvent}\,(\textit{b}) the linear resolvent operator of the fully developed cylinder wake by plotting its maximum energy gain for a range of temporal frequencies $\omega$.
We observe a single resonant peak (marked by a blue dashed line) at $\omega_r=1.04$, as reported in \cite{symon2018non}.
Finally, we further characterize both the DNS and resolvent operator by plotting 
\textit{i}) the spatial distribution of the kinetic energy of the first harmonic mode from DNS in Fig.~\ref{fig:dnsresolvent}\,(\textit{c}); and
\textit{ii}) the spatial distribution of the kinetic energy of the leading resolvent response mode at frequency $\omega_1$ in Fig.~\ref{fig:dnsresolvent}\,(\textit{d}).
Although there is reasonable agreement between the two kinetic energy distributions in panels (\textit{c}) and (\textit{d}), we observe some differences.
Perhaps the most obvious is that the kinetic energy of the leading resolvent response mode in panel (\textit{d}) is too energetic in the far wake when compared to the first harmonic mode from DNS in panel (\textit{c}). \textcolor{black}{Similar discrepancies have also been observed for other spatially-developing flows, e.g.~jets \citep{schmidt2018spectral} and airfoils \citep{Yeh19,Symon19}.}

%\textcolor{black}{We have shown that resolvent analysis for the two-dimensional cylinder flow reveals a dominant linear mechanism occurring at the frequency that matches the vortex shedding. However, there still exist discrepancies between this linear prediction and the actual flow field, e.g.~the streamwise dissipation of the vortex shedding, higher harmonic frequencies. The main purpose of this study is therefore to investigate the extent to which resolvent analysis can correctly model energy transfer mechanisms and to recover energy transfer pathways across frequencies in the true flow. For a better understanding of nonlinearity, we quantify the contribution of cross-frequency interactions to the energy balance for each harmonics in the actual flow field, which is essential to improve modeling and simulation of nonlinear fluid flows.}

\section{Energy balance framework} \label{sec:methods}
In this section we consider the energy balance achieved \textit{i}) by the cylinder flow overall; and \textit{ii}) by each harmonic mode $\hat{\textbf{\textit{u}}}\hspace{0mm}^{(n)}$ from the harmonic decomposition introduced in Sec.~\ref{sec:III.A}.

\subsection{Harmonic decomposition}\label{sec:III.A}
Motivated by the results of Fig.~\ref{fig:dnsresolvent}\,(\textit{a})---where we observed that the energy of the DNS velocity field is concentrated in a small number of discrete frequencies---\textcolor{black}{we now only consider frequencies that are harmonics of the periodic vortex-shedding frequency. A harmonic expansion is therefore introduced for the fluctuating velocity field $\textbf{\textit{u}}'$:}
\begin{align} \label{equ:fouriersummation}
  \textbf{\textit{u}}'&= \sum_{n\in\mathbb{Z}^{+}} \hat{\textbf{\textit{u}}}\hspace{0mm}^{(n)} e^{j\omega_n t}+\text{c.c.},
\end{align}
and similarly for $p'$ and $\textbf{\textit{f}}'$, where $n\in\mathbb{Z}^{+}$ are the positive integers and ($\text{c.c.}$) denotes the complex conjugate.
\textcolor{black}{Note that considering integer multiples of the vortex shedding frequency is not an \textit{a priori} restriction of the energy balance and energy transfer analyses presented in this study. This decomposition will, however, serve to significantly simplify the analysis when we consider the physical mechanisms by which energy is exchanged across temporal frequencies.}
We again note that more than 99.9\,\% of the total fluctuation kinetic energy is contained in the first three harmonic frequencies (see Sec.~\ref{sec:II.C} and Fig.~\ref{fig:dnsresolvent}\,(\textit{a})). \textcolor{black}{Therefore we consider a truncation to only three modes in the harmonic decomposition \eqref{equ:fouriersummation} for all results (i.e.~$n\leq 3$).
}

\subsection{Energy balance for the flow overall} \label{sec:3.1}
We can form an evolution equation for the total kinetic energy of the fluctuations by taking the kinetic-energy inner product of the perturbation equations Eq.~\eqref{equ:pertur} with $\textbf{\textit{u}}'$:
\begin{equation} \label{equ:innerp}
  \langle \textbf{\textit{u}}', \partial_t\textbf{\textit{u}}'\rangle=\langle \textbf{\textit{u}}',L\textbf{\textit{u}}'-\nabla\textit{p}'\rangle+\langle \textbf{\textit{u}}',\textbf{\textit{f}}'\rangle.
\end{equation}
Expanding each term in Eq.~\eqref{equ:innerp} using tensor notation and averaging in time we arrive at
\begin{equation} \label{equ:fourterms}
  \begin{aligned} 
    \dfrac{d\overline{E}}{dt}
    &=
    \underbrace{\int -\overline{u_i'u_j'}\dfrac{\partial U_i}{\partial x_j}\ d\Omega}_{P}
    \ + \ \underbrace{\Rey^{-1}\int -\overline{\dfrac{\partial u_i'}{\partial x_j}\dfrac{\partial u_i'}{\partial x_j}}\ d\Omega}_{D} \\
    &+ \underbrace{\int(-\dfrac{1}{2}\overline{u_i'u_i'}U_j) \cdot n_j \ d\Gamma_\textrm{out}}_{M} 
    \ + \ \underbrace{\int (-\dfrac{1}{2} \overline{u_i'u_i'u_j'}) \cdot n_j \ d\Gamma_\textrm{out}}_{N},
  \end{aligned}
\end{equation}
Equation \eqref{equ:fourterms} is the Reynolds-Orr equation with additional terms $M$ and $N$ to account for any fluxes of energy out of the domain $\Omega$.
These additional terms appear as line integrals across the domain's outlet $\Gamma_\textrm{out}$ after using Gauss' theorem, the boundary conditions, and the divergence-free condition of Eq.~\eqref{equ:pertur}.
Note that the pressure field has been eliminated due to the zero-stress boundary condition at the outlet, $(-u_i'p'\delta_{ij}+\Rey^{-1}u_i'\partial u_i'/\partial x_j)\cdot n_j=0$, where $n_j$ denotes the outward-pointing normal vector on the boundary.
If vortex shedding is fully developed then $d\overline{E}/dt=0$ and it follows that the four terms in Eq.~\eqref{equ:fourterms} must balance so that their sum is zero.

The four terms in Eq.~\eqref{equ:fourterms} represent energy production ($P$); viscous dissipation ($D$); energy flux out by the mean flow ($M$); and work done by the nonlinear terms ($N$).
(The work done by the nonlinear terms, $N$, can also be interpreted as the flux of energy out of the domain by the fluctuations.)
In general we expect production $P$, which represents the energy extracted from the mean flow by the fluctuations, to be positive.
Viscous dissipation $D$, meanwhile, is always negative.
The energy flux terms $M$ and $N$ represent the energy leaving the domain by linear ($M$) or nonlinear ($N$) mechanisms.
%and are therefore negative in general.
Note that both $M$ and $N$ tend to zero if the outlet boundary is placed infinitely far from the cylinder since any fluctuations will dissipate before reaching the boundary.
In this case Eq.~\eqref{equ:fourterms} simplifies to the Reynolds--Orr equation which states that, when considered over the entire domain, dissipation balances production.

\subsection{Energy balance for each harmonic mode} \label{sec:3.2}
We can derive a similar energy balance for each harmonic mode by substituting the harmonic decomposition Eq.~\eqref{equ:fouriersummation} into Eq.~\eqref{equ:pertur} and (after using the orthogonality of the complex exponentials) taking the inner product with $\hat{\textbf{\textit{u}}}^*$:
%, we arrive at an energy balance equation for each harmonic mode:
\begin{equation} \label{equ:innerharm}
  \langle \hat{\textbf{\textit{u}}}^*, j\omega\hat{\textbf{\textit{u}}} \rangle + \text{c.c.}
  =
  \langle \hat{\textbf{\textit{u}}}^*,L\hat{\textbf{\textit{u}}}-\nabla\hat{\textit{p}}\rangle+\text{c.c.}+\langle \hat{\textbf{\textit{u}}}^*,\hat{\textbf{\textit{f}}}\rangle+\text{c.c.}
\end{equation}
(For simplicity the superscript on $\hat{\textbf{\textit{u}}}$ in Eq.~\eqref{equ:fouriersummation} has been removed.)
Expanding Eq.~\eqref{equ:innerharm} and again using tensor notation we arrive at
\begin{equation} \label{equ:reyorr_fre2}
  \begin{aligned}
    j\omega\hat{E}+\text{c.c.}
    &=
    \underbrace{\int -(\hat{u}^*_i\hat{u}_j+\hat{u}_i\hat{u}^*_j)\dfrac{\partial U_i}{\partial x_j}\ d\Omega}_{\hat{P}(\omega)}
    \ + \ \underbrace{\int (\hat{u}^*_i\hat{f}_i+\hat{u}_i\hat{f}^*_i) \ d\Omega}_{\hat{N}(\omega)}\\
    &+ \ \underbrace{\dfrac{2}{\Rey}\int -\dfrac{\partial \hat{u}^*_i}{\partial x_j}\dfrac{\partial \hat{u}_i}{\partial x_j}\ d\Omega}_{\hat{D}(\omega)}
    \ + \underbrace{\int(-\hat{u}^*_i\hat{u}_iU_j)\cdot n_j\ d\Gamma_\textrm{out}}_{\hat{M}(\omega)}.
  \end{aligned}
\end{equation}
The left-hand side of Eq.~\eqref{equ:reyorr_fre2} satisfies $j\omega\hat{E}+\text{c.c.}=0$ for fully developed vortex shedding, indicating that each harmonic mode neither gains nor loses energy over one cycle.
Therefore similar to the global energy balance in Eq.~\eqref{equ:fourterms}, there also exists a balance for each harmonic mode across the four terms $\hat{P}(\omega)$, $\hat{D}(\omega)$, $\hat{M}(\omega)$, and $\hat{N}(\omega)$.
Note that, due to the summation over complex conjugate pairs, all terms in Eq.~\eqref{equ:reyorr_fre2} are real-valued.

%The forcing $\hat{\textbf{\textit{f}}}$ in Eq.~\eqref{equ:reyorr_fre2} could either represent the true nonlinear forcing from DNS or be replaced by the leading input resolvent mode given by Eq.~\eqref{equ:resolvent_opti}.
\textcolor{black}{For the nonlinear flow, the input  $\hat{\textbf{\textit{f}}}$ in Eq.~\eqref{equ:reyorr_fre2} represents the true nonlinear forcing at each harmonic frequency.}
The term $\hat{N}(\omega_n)$ represents the work done by the nonlinear forcing on harmonic mode $n$ and can be positive or negative.
If $\hat{N}(\omega_n)$ is positive (negative) then the nonlinear terms give energy to (take energy from) harmonic mode $n$.
\textcolor{black}{The energy transfer due to the nonlinear forcing is compared to that predicted by the optimal resolvent forcing mode, which excites the most linearly amplified response, in Sec.~\ref{sec:result1}.}
It is important to note that, due to the unitary property of the Fourier transform, we may link the energy balance at each frequency $\omega_n$ to the energy balance for the flow overall.
For production, for example:
\begin{equation}
  P = \sum_n \hat{P}(\omega_n),
\end{equation}
and similarly for $D$, $M$, and $N$.
(Note that this will hold only approximately due to the truncation to three modes of the harmonic decomposition.)

\section{Energy balance in the cylinder wake}\label{sec:result1}
We now compare the energy balance achieved across production, dissipation, and nonlinear transfer for DNS and resolvent analysis.
Specifically we will consider the energy balance
\textit{i}) for each harmonic mode;
\textit{ii}) for the flow as a whole; and
\textit{iii}) for the nonlinear transfer between modes.
It will be convenient to define a linear dissipation term, $\hat{D}_e(\omega)=\hat{D}(\omega)+\hat{M}(\omega)$ for each harmonic mode and $D_e=D+M$ for the flow as a whole, to denote the total effective dissipation due to linear mechanisms.
The energy balance across production, linear dissipation, and nonlinear transfer for DNS and resolvent analysis is plotted in Figs.~\ref{fig:energybalance_comparison} and \ref{fig:energybalance_dns}, which show the same information but in different ways.
In Fig.~\ref{fig:energybalance_comparison} the energy balance is arranged by harmonic mode, each row representing a single harmonic.
In Fig.~\ref{fig:energybalance_dns} the energy balance is instead arranged by physical mechanism: the first row for production, $P$; the second row for the total linear dissipation, $D_e$; and the third row for nonlinear transfer, $N$.
In both figures the energy balance is shown for DNS in panel (\textit{a}) and for resolvent analysis in panel (\textit{b}).
\subsection{Energy balance for the DNS} \label{sec:balDNS}
\begin{figure}
  \centerline{\includegraphics[width=0.925\textwidth]{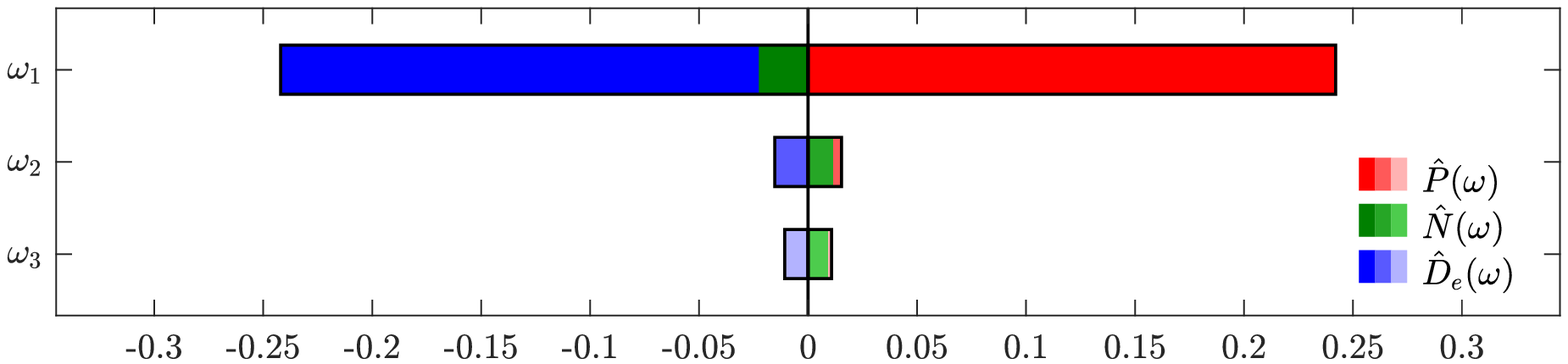}
    \llap{\parbox[b]{13.1in}{(a)\\\rule{0ex}{1.4in}}}}
  \centerline{\includegraphics[width=0.925\textwidth]{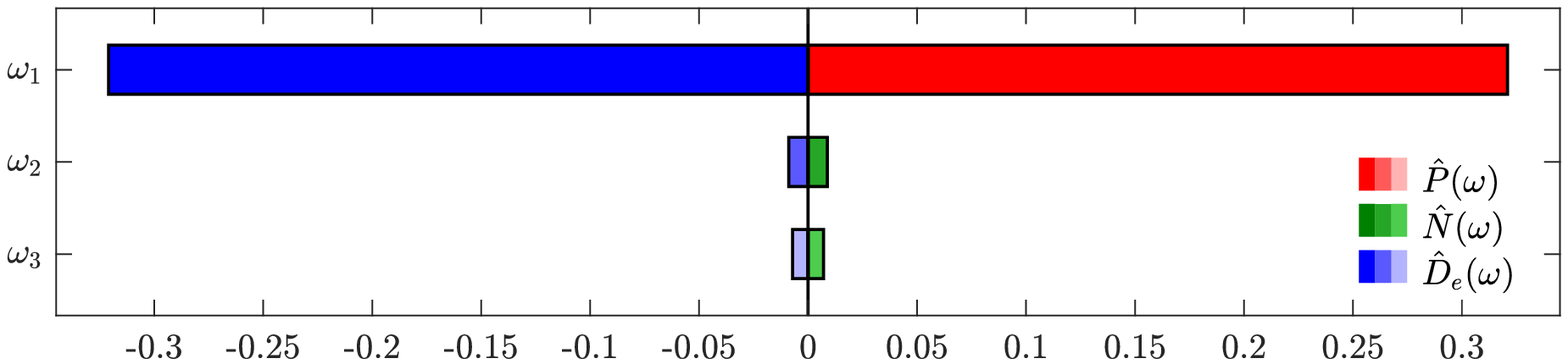}
    \llap{\parbox[b]{13.1in}{(b)\\\rule{0ex}{1.4in}}}}
  \caption{Harmonic energy balance at the first three harmonic frequencies for $(\text{a})$ DNS and $(\text{b})$ resolvent analysis across production, linear dissipation, and nonlinear energy transfer, denoted as $\hat{P}(\omega)$, $\hat{D}_e(\omega)$, and $\hat{N}(\omega)$ respectively.}
  \label{fig:energybalance_comparison}
\end{figure}

\begin{figure}
	\centerline{\includegraphics[width=0.925\textwidth]{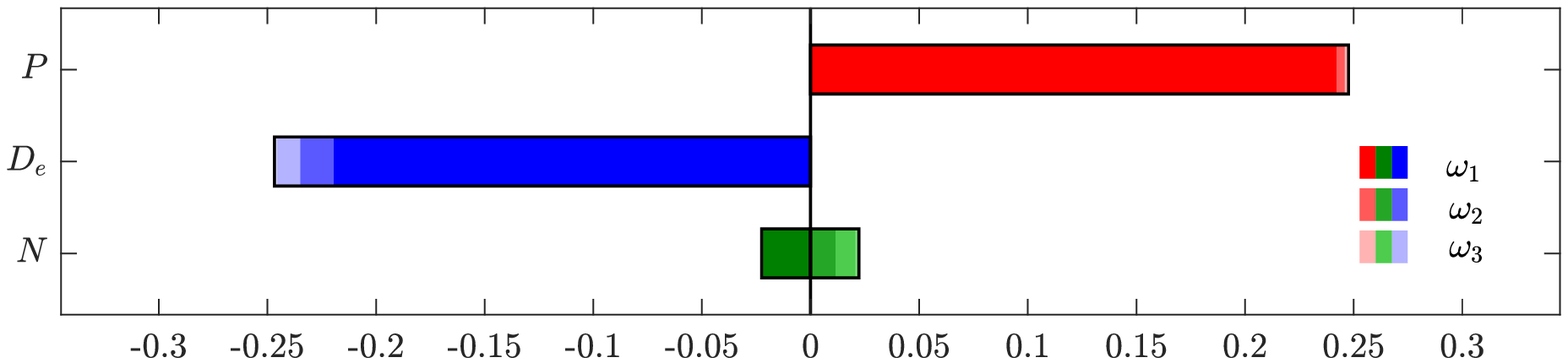}
	\llap{\parbox[b]{13.1in}{(a)\\\rule{0ex}{1.4in}}}}
	\centerline{\includegraphics[width=0.925\textwidth]{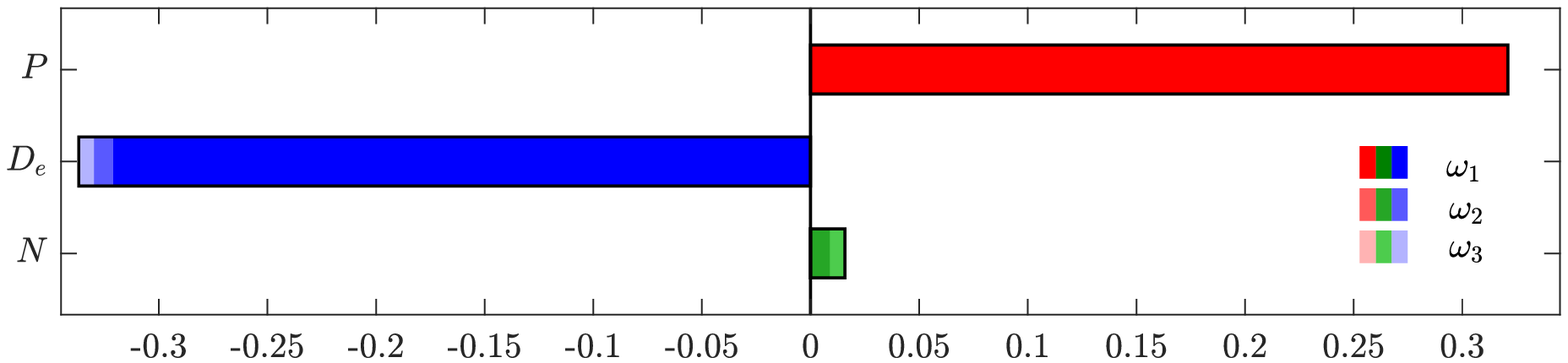}
	\llap{\parbox[b]{13.1in}{(b)\\\rule{0ex}{1.4in}}}}
	\caption{Energy balance for (\text{a}) DNS and (\text{b}) resolvent analysis (summing over the first three harmonic frequencies) across production, linear dissipation, and nonlinear energy transfer, which are denoted as $P$, $D_e$, and $N$, respectively.}
	\label{fig:energybalance_dns}
\end{figure}

Let us start with the DNS data.
The first observation is that the vast majority of both production and dissipation is achieved by the first harmonic mode, $\omega_1$.
We also note that, for each harmonic mode, there exists an energy balance between production $\hat{P}(\omega_n)$, dissipation $\hat{D}_e(\omega_n)$, and nonlinear transfer $\hat{N}(\omega_n)$:
\begin{equation} \label{eq:powerBalMode}
  \hat{P}(\omega_n) + \hat{D}_e(\omega_n) + \hat{N}(\omega_n)= 0.
\end{equation}
This is most clearly seen in Fig.~\ref{fig:energybalance_comparison}\,(\textit{a}).
For the first harmonic mode, production exceeds dissipation ($\hat{P}(\omega_1)>\hat{D}_e(\omega_1)$) and the difference between them is balanced by a negative nonlinear transfer ($\hat{N}(\omega_1)<0$).
For the second and third harmonics the inverse is true: for both modes, dissipation exceeds production ($\hat{D}_e(\omega_2)>\hat{P}(\omega_2)$, $\hat{D}_e(\omega_3)>\hat{P}(\omega_3)$) and for each the difference between them is balanced by a positive nonlinear transfer ($\hat{N}(\omega_2)>0$, $\hat{N}(\omega_3)>0$).
We also observe a similar balance between production, dissipation, and nonlinear transfer across all modes in aggregate:
\begin{equation} \label{eq:powerBalBox}
  \sum_n ( \hat{P}(\omega_n) + \hat{D}_e(\omega_n) + \hat{N}(\omega_n)) = 0.
\end{equation}
This second balance follows naturally from the first by summing Eq.~\eqref{eq:powerBalMode} over the three harmonic modes.
%The energy of the overall flow is the sum of the energy in each mode, and we have seen that a balance between production, dissipation and nonlinear transfer is achieved for each mode \eqref{eq:powerBalMode}.
It therefore follows that there exists a similar balance between production, dissipation, and nonlinear transfer for the flow as a whole.
More importantly, we identify an energy balance that is achieved between production $P$ and dissipation $D_e$ for the flow as a whole:
\begin{equation} \label{eq:powerBalPD}
  \underbrace{\sum_n \hat{P}(\omega_n)}_{P} + \underbrace{\sum_n \hat{D}_e(\omega_n)}_{D_e} \approx 0,
\end{equation}
\textcolor{black}{which only consists of linear mechanisms, as revealed by the Reynolds-Orr equation derived for spatially localized or spatially periodic fluctuations \citep{schmid2002stability}.} Note that it naturally results from another energy balance achieved by the nonlinear transfer terms only:
\begin{equation} \label{eq:nonlinBalBox}
  N = \sum_n \hat{N}(\omega_n) \approx 0.
\end{equation}
\textcolor{black}{This balance is explained by the conservative nature of the nonlinearity---that is it neither creates nor destroys energy when integrated over the entire domain (i.e.~$N=0$), which can be most clearly seen from the Reynolds-Orr equation \eqref{equ:fourterms}, where the nonlinearity accounts only for any fluctuation energy leaving the domain. With a sufficiently large computational domain, all fluctuations would eventually vanish at the boundary and thus the energy flux out of the domain due to the nonlinear terms would be negligible. This is approximately true for the cylinder flow, as shown by the third row (in green) of Fig.~\ref{fig:energybalance_dns}\,(\textit{a}).}
%This balance is most clearly seen in Fig.~\ref{fig:energybalance_dns}\,(\textit{a}) and Eq.~\eqref{equ:fourterms}. It implies that the energy flux out of the domain due to the nonlinear terms is negligible and therefore that the nonlinear transfer terms are conservative---that is they neither create nor destroy energy when integrated over the entire flow.

It is instructive now to consider the manner in which the nonlinear balance in Eq.~\eqref{eq:nonlinBalBox} is achieved in DNS.
From Fig.~\ref{fig:energybalance_dns}\,(\textit{a}) we observe that the nonlinear transfer is negative for the first harmonic ($\hat{N}(\omega_1)<0$) and positive for the higher harmonics ($\hat{N}(\omega_2)>0$, $\hat{N}(\omega_3)>0$).
This implies that the first harmonic loses energy by nonlinear transfer, which is balanced by positive nonlinear transfer for the remaining harmonics.
Together the nonlinear transfer terms therefore act as an inter-mode mediator, taking energy from modes that produce more energy than they dissipate and giving it to modes that dissipate more energy than they produce.


\subsection{Energy balance for resolvent analysis} \label{sec:balRes}
We now consider the extent to which the energy balances established in Sec.~\ref{sec:balDNS} for the DNS are respected by resolvent analysis.
We stress here the assumption implicit in the resolvent analysis is that each harmonic mode is forced by its leading resolvent forcing mode at its corresponding frequency $\omega_1$, $\omega_2$ or $\omega_3$.
Note that, with \textcolor{black}{unit} forcing input, the kinetic energy of each harmonic response mode is set by the corresponding singular value squared $\sigma_1^2(\omega_n)$. 
We have therefore selected forcing mode amplitudes such that the total kinetic energy of each harmonic response is the same as that in the DNS.
By doing so we first note in Fig.~\ref{fig:energybalance_dns} that the energy production $\hat{P}(\omega_1)$ is actually slightly higher for the resolvent case than it is for the DNS. It is expected that the amount of production will be different since the DNS and resolvent mode shapes in Figs.~\ref{fig:dnsresolvent}\,(\textit{c}) and (\textit{d}) are different.

For each harmonic mode, the energy balance in Eq.~\eqref{eq:powerBalMode} is achieved by resolvent analysis, as depicted in Fig.~\ref{fig:energybalance_comparison}\,(\textit{b}).
This balance ensures that each harmonic mode, as modeled by the resolvent, neither gains nor loses energy over a cycle.
This balance for each harmonic mode in turn ensures that there also exists a similar balance over all modes, i.e.~that the second balance in Eq.~\eqref{eq:powerBalBox} is also satisfied by resolvent analysis.
In particular, we see in Fig.~\ref{fig:energybalance_comparison}\,(\textit{b}) that the energy balance achieved between production and dissipation now only applies for the first harmonic:
\begin{equation}
    \hat{P}(\omega_1) + \hat{D}_e(\omega_1)\approx 0,
\end{equation}
which can be concluded from the fact that $\hat{N}(\omega_1)\approx 0$ in Fig.~\ref{fig:energybalance_comparison}\,(\textit{b}). \textcolor{black}{These results are consistent with those observed in wall-bounded shear flows \citep{symon2020energy}.}
We come now to a key difference between the DNS and resolvent analysis: the energy balance for the nonlinear transfer terms $\hat{N}(\omega_n)$.
\textcolor{black}{Unlike the DNS, the resolvent analysis considers only linear mechanisms that are most excited by external disturbances. This leads to a violation of the nonlinear balance in Eq.~\eqref{eq:nonlinBalBox} when considering the flow as a whole, as shown in Fig.~\ref{fig:energybalance_dns}\,(\textit{b}):}
\begin{equation} 
  N = \sum_n \hat{N}(\omega_n) \neq 0.
\end{equation}
\textcolor{black}{The resolvent-based predictions are not expected to satisfy the nonlinear balance in Eq.~\eqref{eq:nonlinBalBox} since the resolvent analysis considers only linear amplification mechanisms and does not consider energy transfer between harmonic modes. We therefore observe that the energy balance among these harmonics does not satisfy the constraints placed on them by the Navier--Stokes equations. The implication is that, when each harmonic mode is assumed to be forced by its leading forcing mode, additional constraints or information is required to take into account the effect of the nonlinear forcing, e.g.~a time-varying base flow \citep{padovan_otto_rowley_2020}, an extended resolvent operator \citep{marquet2020extended} or a harmonic balance model \citep{rigas2020non}.}

\subsection{Spatial distribution of the energy balance} \label{sec:balSpace}
Having looked at the balance of production, dissipation, and nonlinear transfer over the domain as a whole in Secs.~\ref{sec:balDNS} \& \ref{sec:balRes}, we now consider the distribution of these terms in physical space.
By doing so we identify the regions of the flow most responsible for production, dissipation, and nonlinear transfer for each of the three harmonic modes.

Let us first consider the spatial distributions for DNS, which are plotted in Fig.~\ref{tab:dnscomponents}.
Each row represents a harmonic frequency (with the final row representing their aggregate) and each column represents a single physical mechanism.
%\textcolor{black}{\st{Although linear dissipation $\hat{D}_e(\omega)$ is generally considered as a whole, we here plot the contributions of viscous dissipation $\hat{D}(\omega)$ and energy flux out $\hat{M}(\omega)$ separately using the top and bottom halves of each panel.}}
We note from the color scale that, consistent with Figs.~\ref{fig:energybalance_comparison} \& \ref{fig:energybalance_dns}, the two dominant terms are production and dissipation for the first harmonic, $\hat{P}(\omega_1)$ and $\hat{D}_e(\omega_1)$.
This is also clearly seen by comparing the production for the first harmonic with the overall production; and the dissipation for the first harmonic with the overall dissipation.
We also see that, like production and dissipation, the nonlinear transfer terms are significantly larger for the first harmonic than any other. 
\textcolor{black}{
In particular, the nonlinear transfer term for the first harmonic $\hat{N}(\omega_1)$ is negative along the centreline which coincides with the maximum of the energy production term $\hat{P}(\omega_1)$. In other words, the nonlinear term $\hat{N}(\omega_1)$ directly extracts energy from where it is generated and transfers it to the second harmonic (i.e.~$\hat{N}(\omega_2)$ is positive along the centerline). Thus, the first harmonic mode in the true flow is substantially shorter in the streamwise direction than the optimal resolvent mode (for which nonlinear energy transfer is neglected, see Fig.~\ref{fig:dnsresolvent}). Another significant observation is the spatial consistency observed between the dissipation and nonlinear transfer for the third harmonic frequency, which implies that the energy transferred into the third harmonic is dissipated locally.
}

\newcolumntype{C}{>{\centering\arraybackslash}m{0.2725\textwidth}}
\newcolumntype{a}{>{\centering\arraybackslash}m{0.05\textwidth}}
\aboverulesep=0ex
\belowrulesep=0ex
\begin{figure*}
	\begin{tabular}{m{7mm} | C | C | C a}%l*3{C}@{}
		%\toprule
		$\ $ & $\hat{P}(\omega)$ & \textcolor{black}{$\hat{D}_e(\omega)$} & $\hat{N}(\omega)$&$\ $ \\[2.5pt]
		\midrule
		&&&&\\[-0.8em]
        \llap{\parbox[b]{0in}{$\omega_1$\\\rule{0ex}{0.2in}}}& \includegraphics[width=0.2725\textwidth]{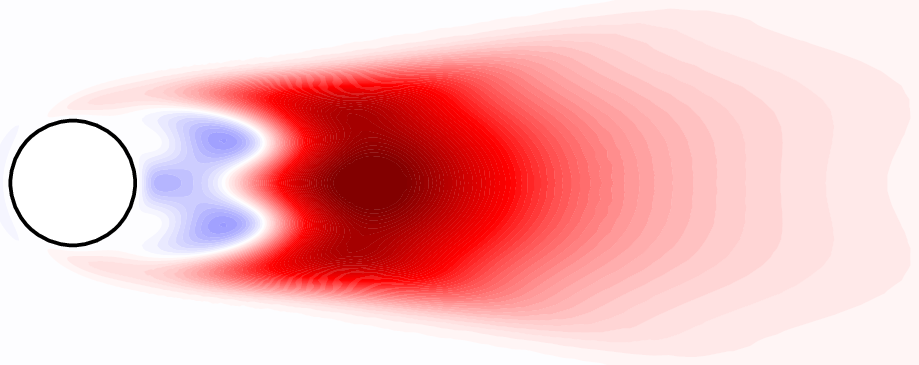}&
        \includegraphics[width=0.2725\textwidth]{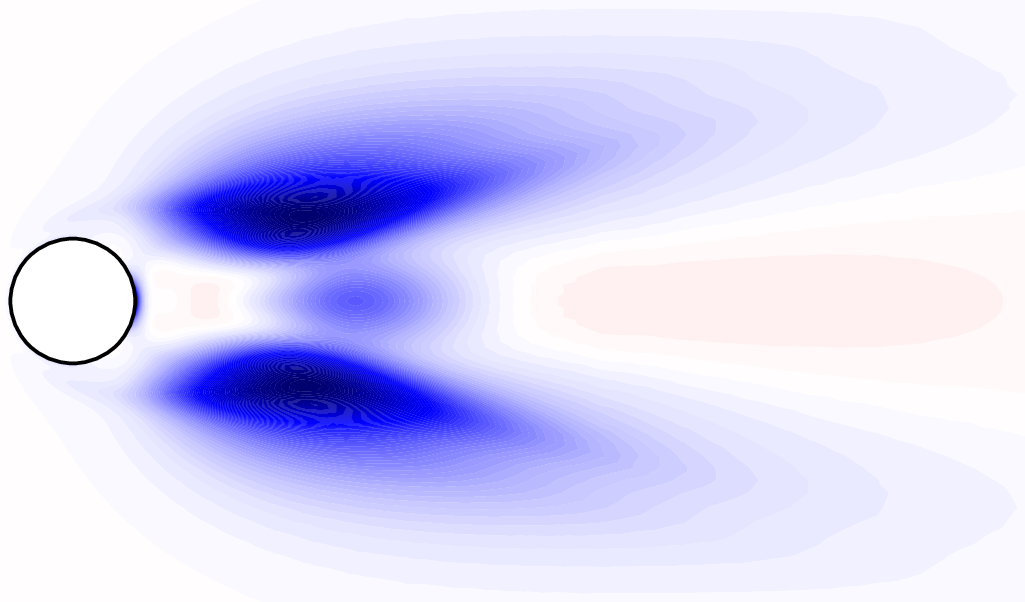}& \includegraphics[width=0.2725\textwidth]{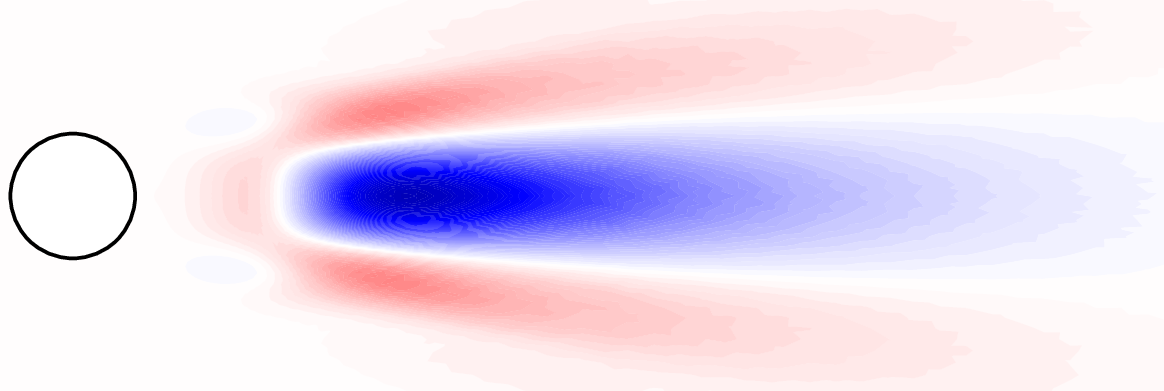}&
		\includegraphics[width=0.05\textwidth]{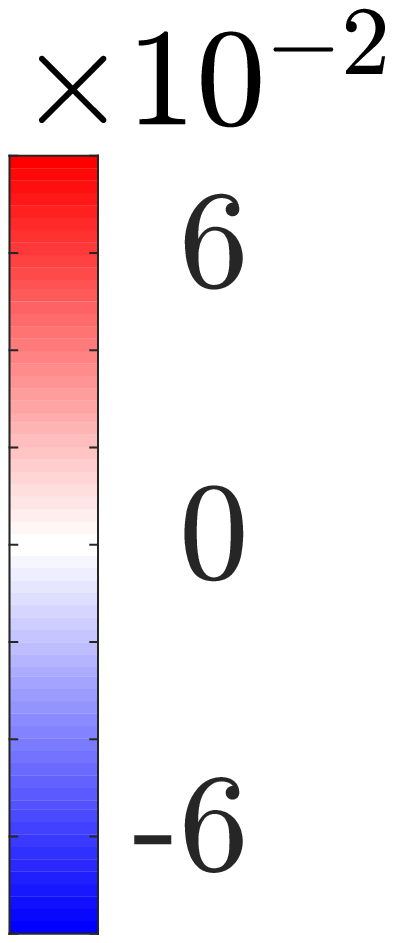}\\[2.5pt]
		\midrule
		&&&&\\[-0.8em]
		\llap{\parbox[b]{0in}{$\omega_2$\\\rule{0ex}{0.2in}}}& \includegraphics[width=0.2725\textwidth]{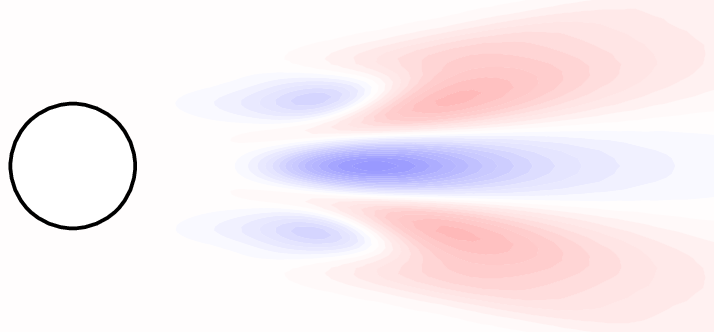} & \includegraphics[width=0.2725\textwidth]{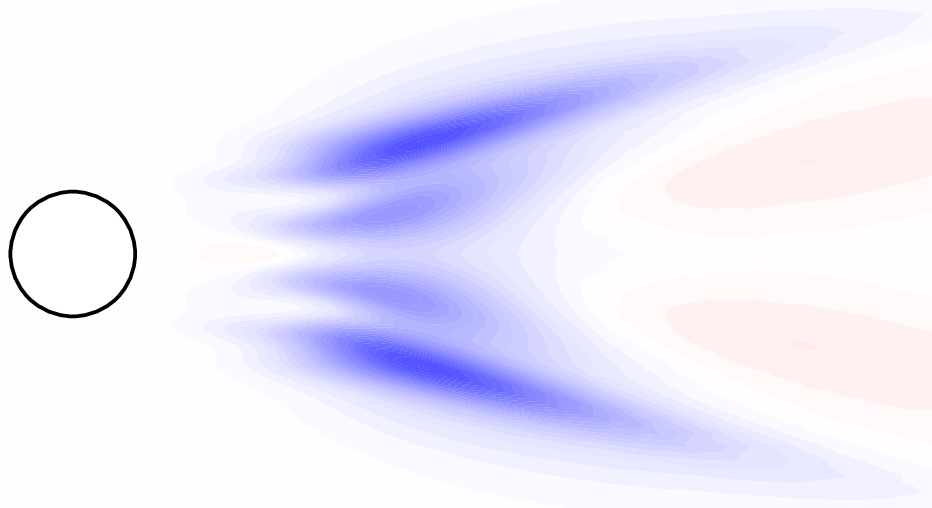}
		& \includegraphics[width=0.2725\textwidth]{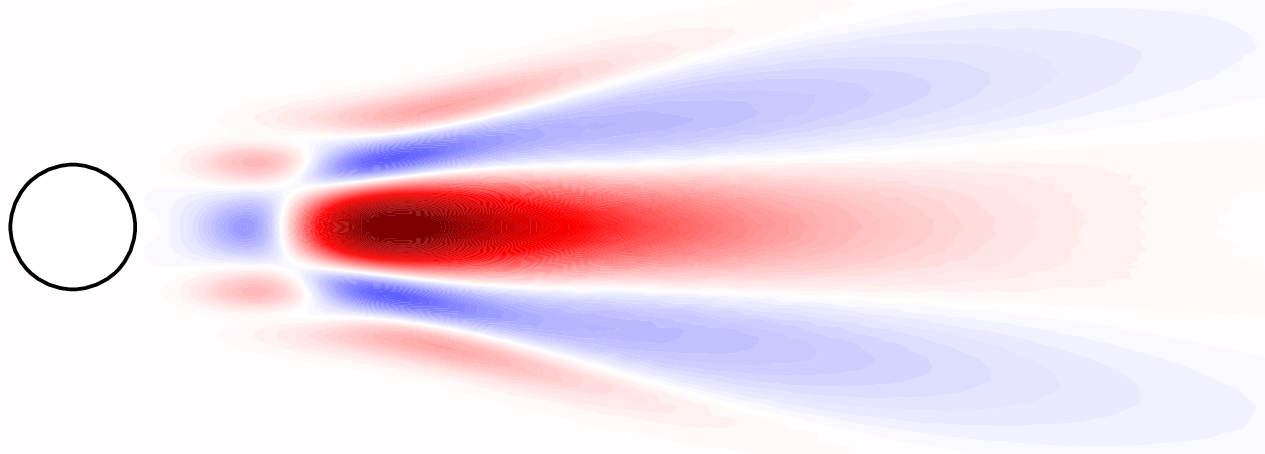}&
		\includegraphics[width=0.05\textwidth]{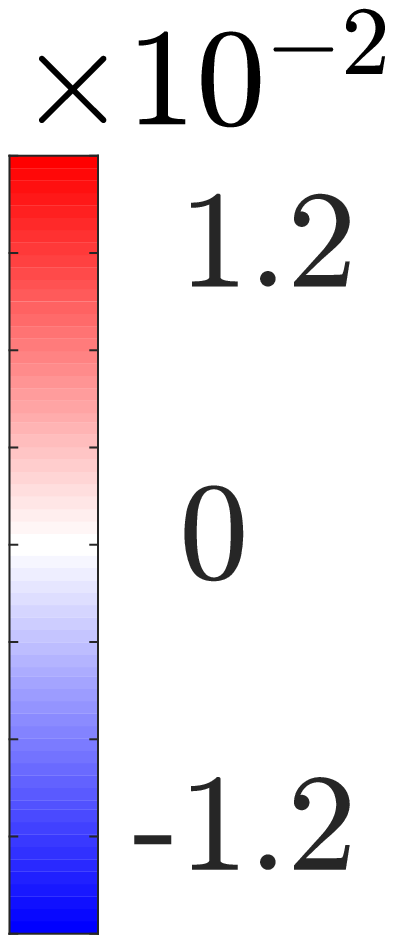}\\[2.5pt]
		\midrule
		&&&&\\[-0.8em]
		\llap{\parbox[b]{0in}{$\omega_3$\\\rule{0ex}{0.2in}}}& \includegraphics[width=0.2725\textwidth]{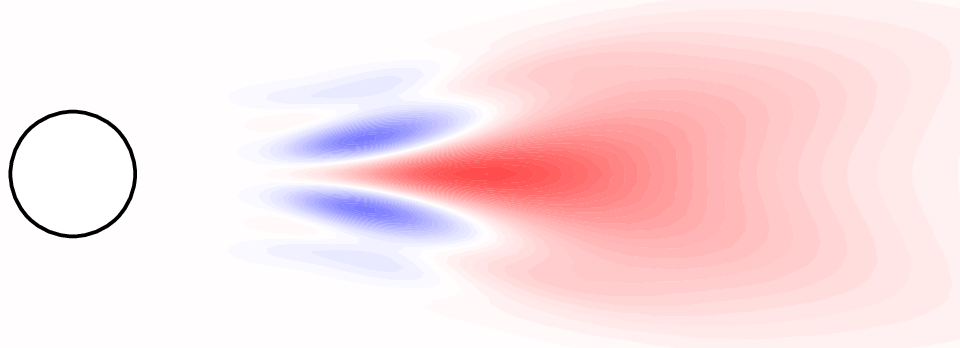} & \includegraphics[width=0.2725\textwidth]{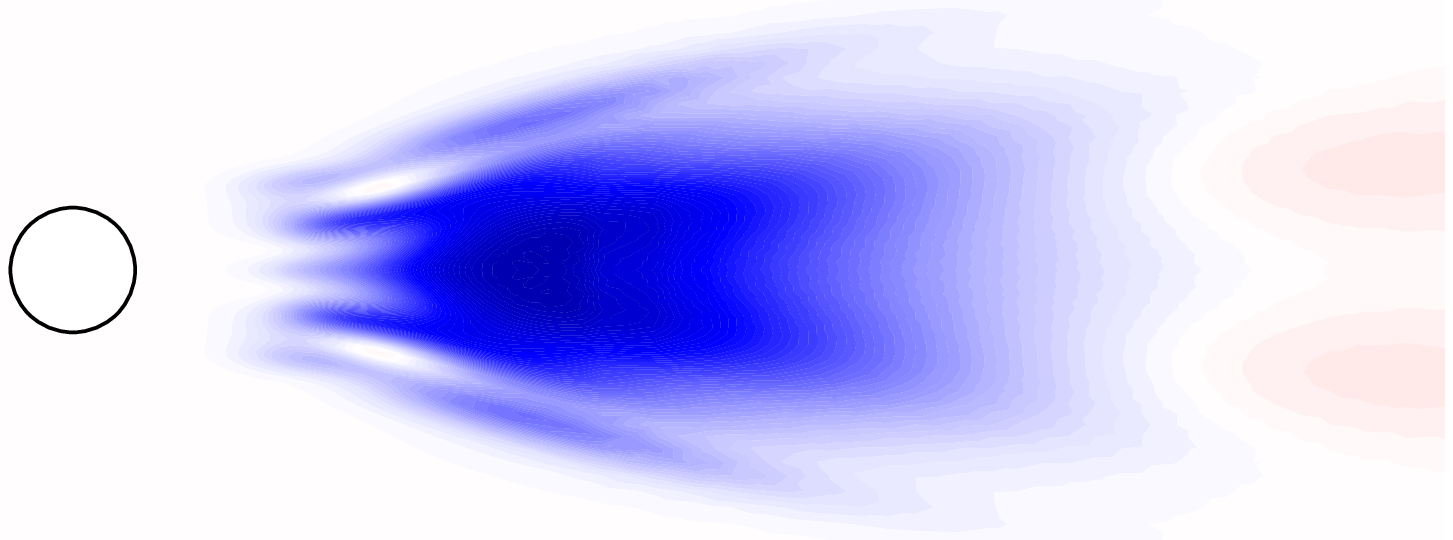}
		& \includegraphics[width=0.2725\textwidth]{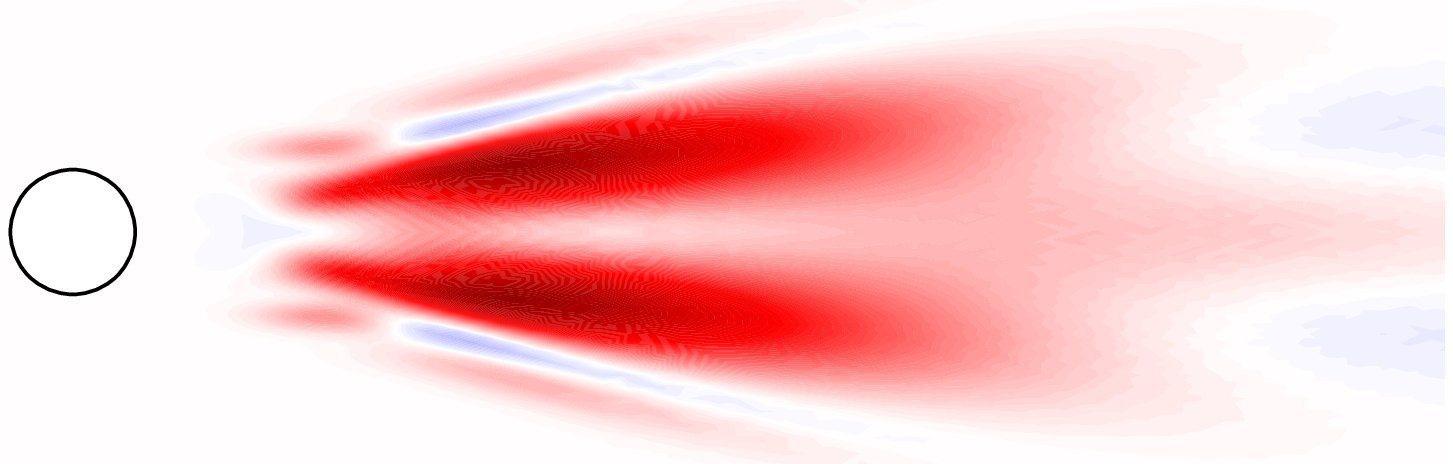}&
		\includegraphics[width=0.05\textwidth]{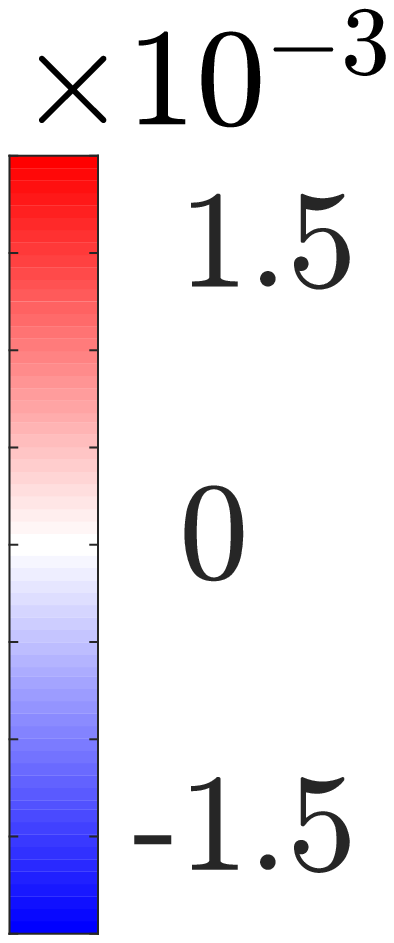}\\[2.5pt]
		\midrule
		&&&&\\[-0.8em]
		%$\cdots$&$\cdots$&$\cdots$&$\cdots$\\
		\llap{\parbox[b]{0in}{DNS\\\rule{0ex}{0.2in}}}& \includegraphics[width=0.2725\textwidth]{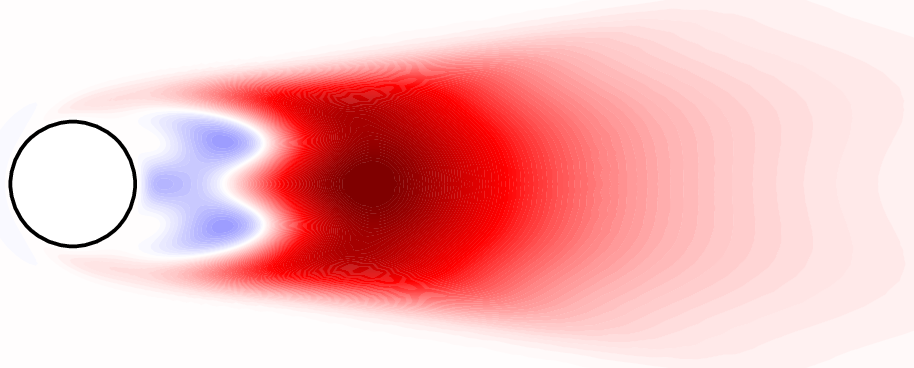} & \includegraphics[width=0.2725\textwidth]{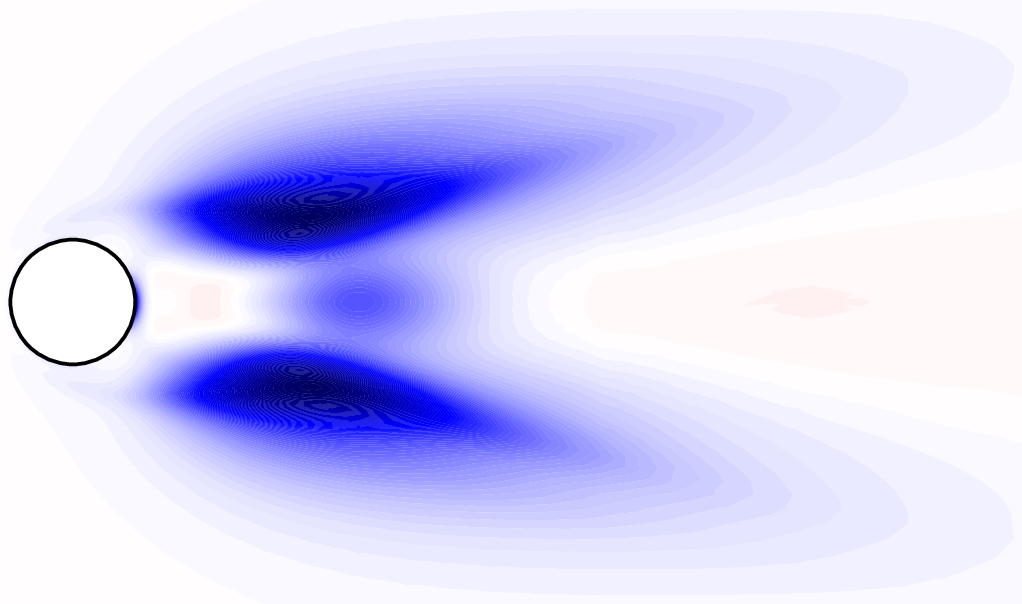} 
		& \includegraphics[width=0.2725\textwidth]{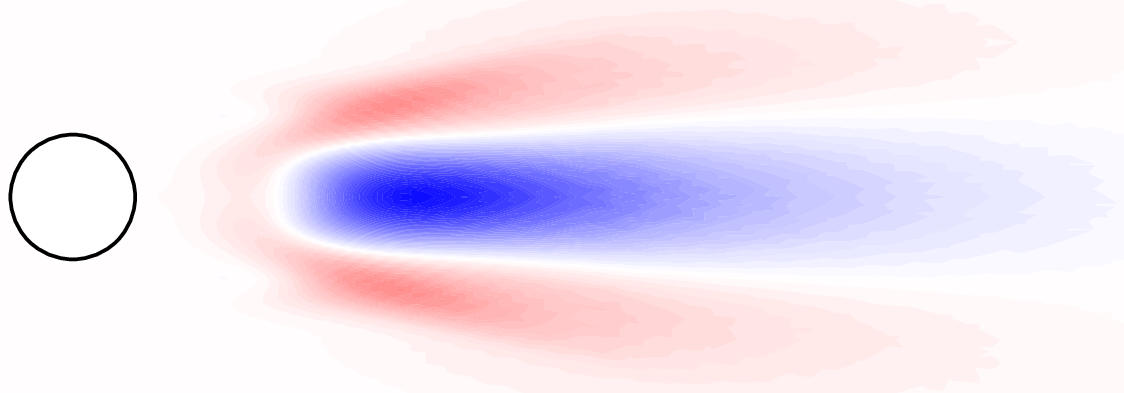}&
		\includegraphics[width=0.05\textwidth]{fig4d.eps}\\[2.5pt]
		\midrule 
	\end{tabular}
	\caption{\textcolor{black}{Spatial distribution of production $\hat{P}(\omega)$, linear energy dissipation $\hat{D}_e(\omega)$, and nonlinear energy transfer $\hat{N}(\omega)$ from DNS for each harmonic frequency. Note the smaller color scales for $\omega_2$ and $\omega_3$.}} \label{tab:dnscomponents}
\end{figure*} 

\begin{figure}
	\begin{tabular}{m{7mm} | C | C | C a}
		%\toprule
		$\ $ & $\hat{P}(\omega)$ & \textcolor{black}{$\hat{D}_e(\omega)$} & $\hat{N}(\omega)$&$\ $ \\[2.5pt]
		\midrule
		&&&&\\[-0.8em]
		\llap{\parbox[b]{0in}{$\omega_1$\\\rule{0ex}{0.2in}}}& \includegraphics[width=0.2725\textwidth]{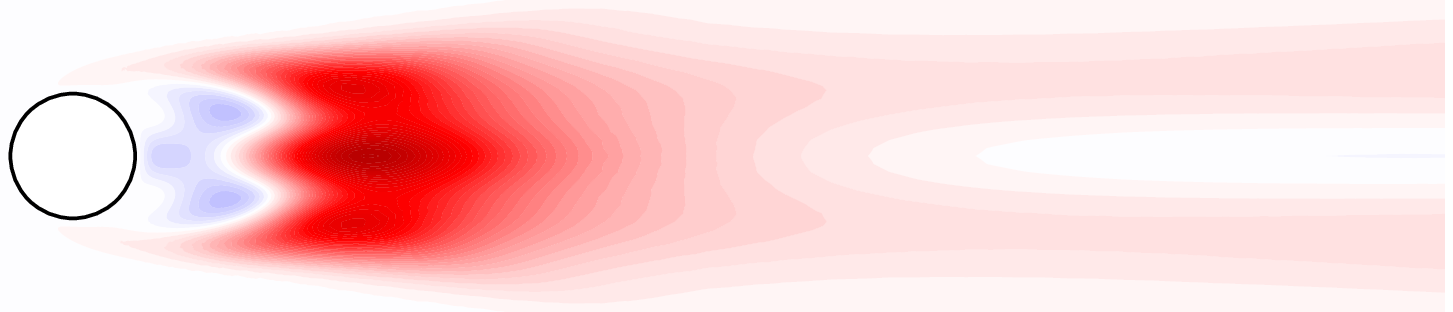} & \includegraphics[width=0.2725\textwidth]{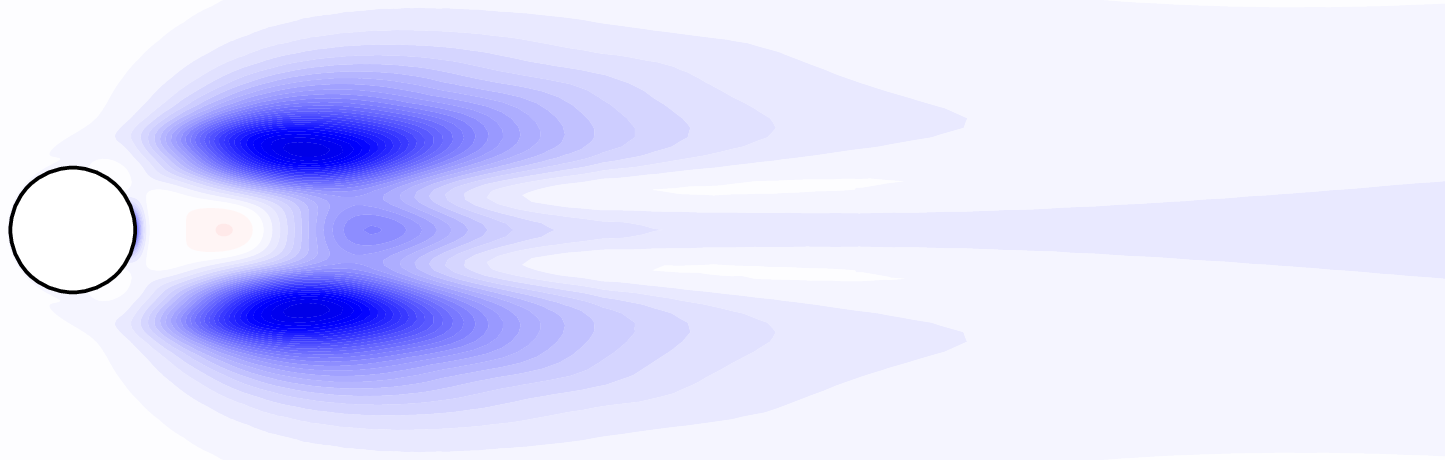}
		& \includegraphics[width=0.2725\textwidth]{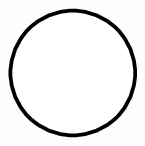} &\includegraphics[width=0.05\textwidth]{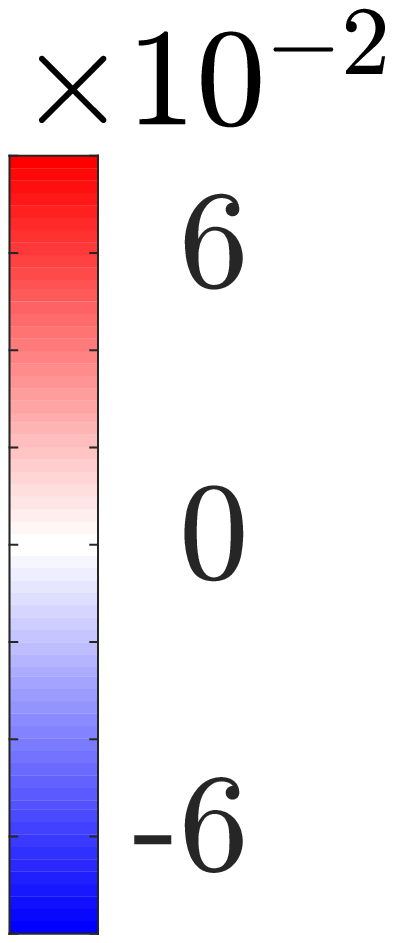}\\[2.5pt]
		\midrule
		&&&&\\[-0.8em]
		\llap{\parbox[b]{0in}{$\omega_2$\\\rule{0ex}{0.2in}}}& \includegraphics[width=0.2725\textwidth]{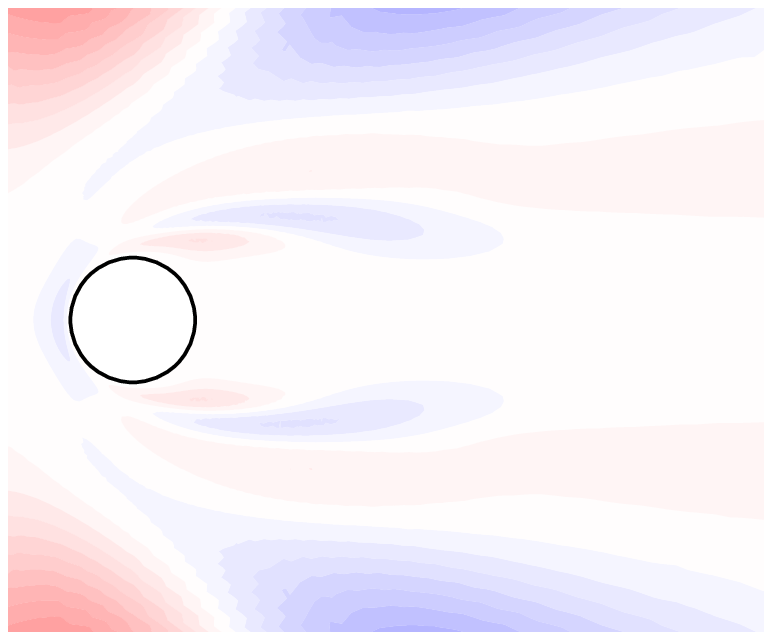} & \includegraphics[width=0.2725\textwidth]{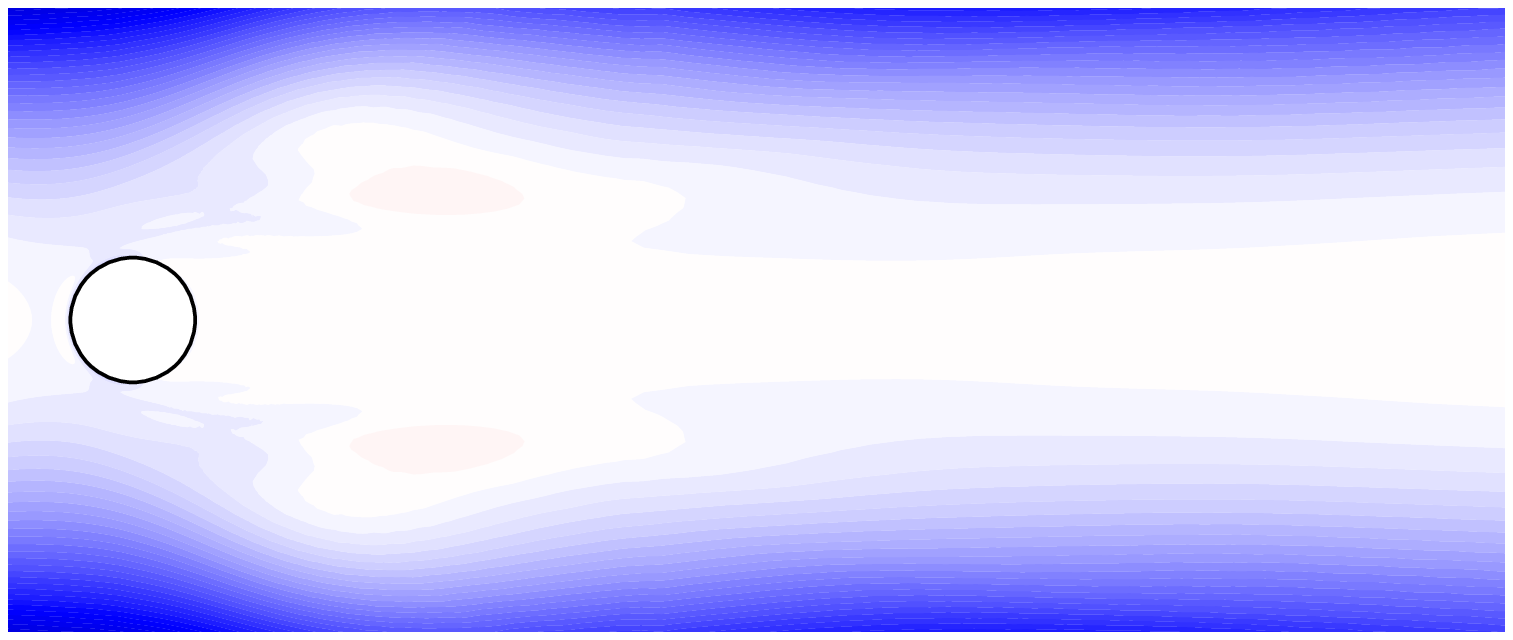}
		& \includegraphics[width=0.2725\textwidth]{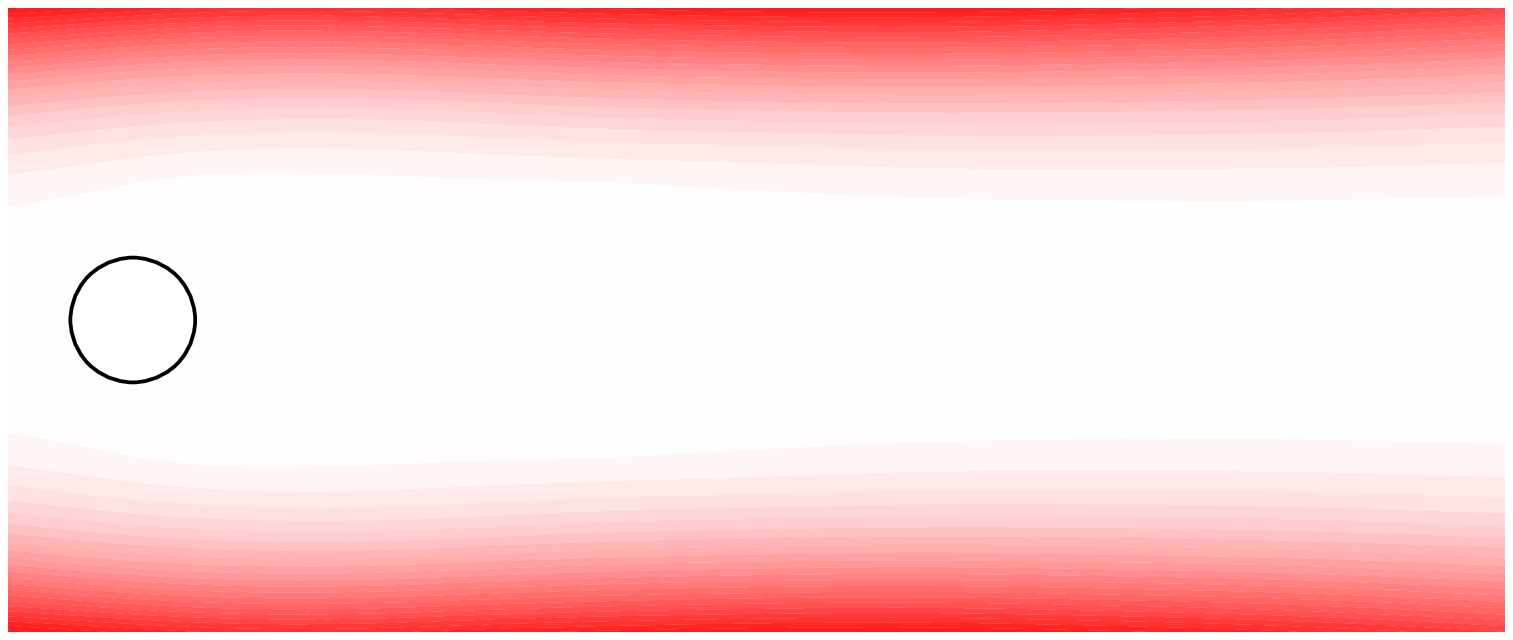}&
		\includegraphics[width=0.05\textwidth]{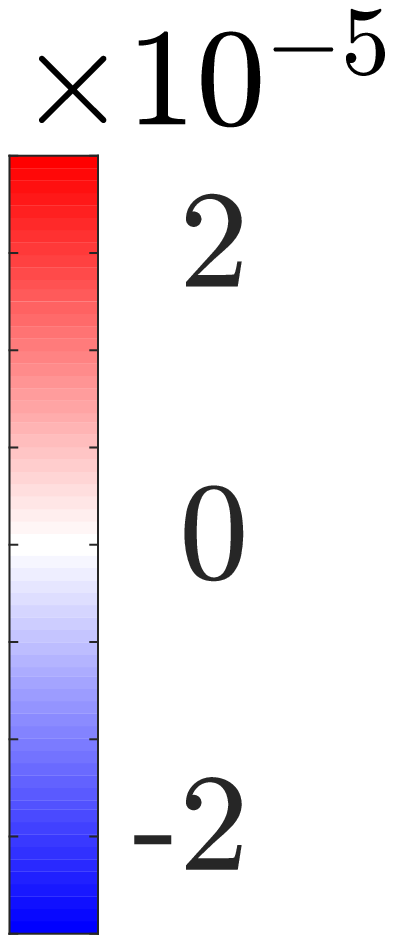}\\[2.5pt] 
		\midrule
		&&&&\\[-0.8em]
		\llap{\parbox[b]{0in}{$\omega_3$\\\rule{0ex}{0.2in}}}& \includegraphics[width=0.2725\textwidth]{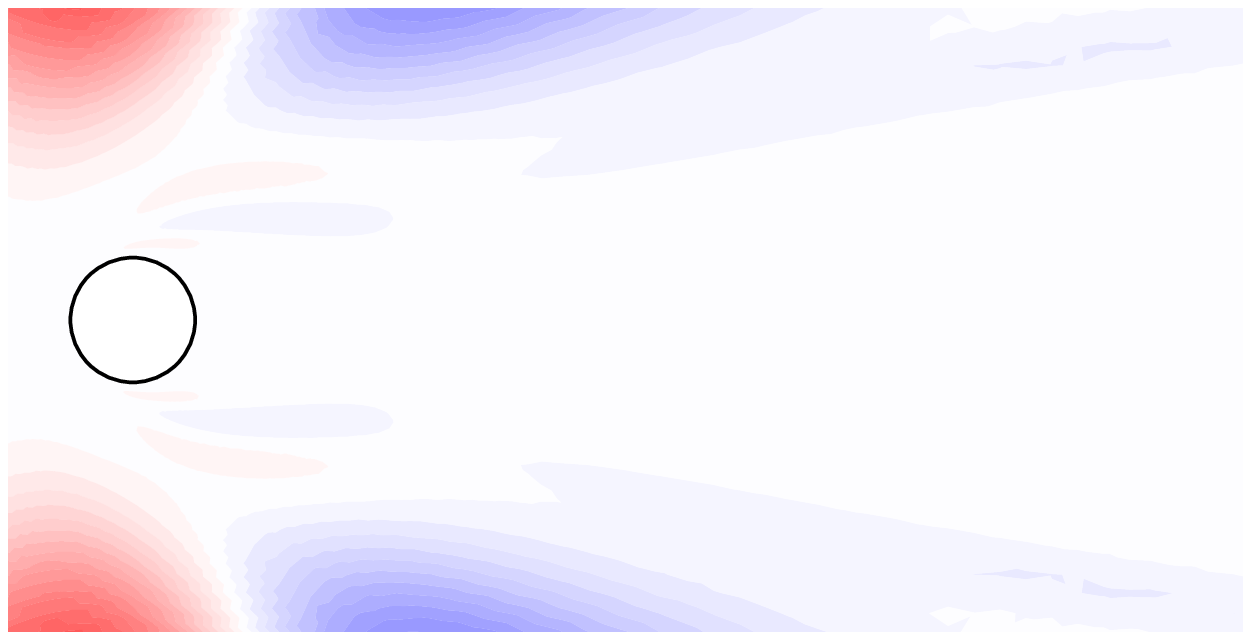} & \includegraphics[width=0.2725\textwidth]{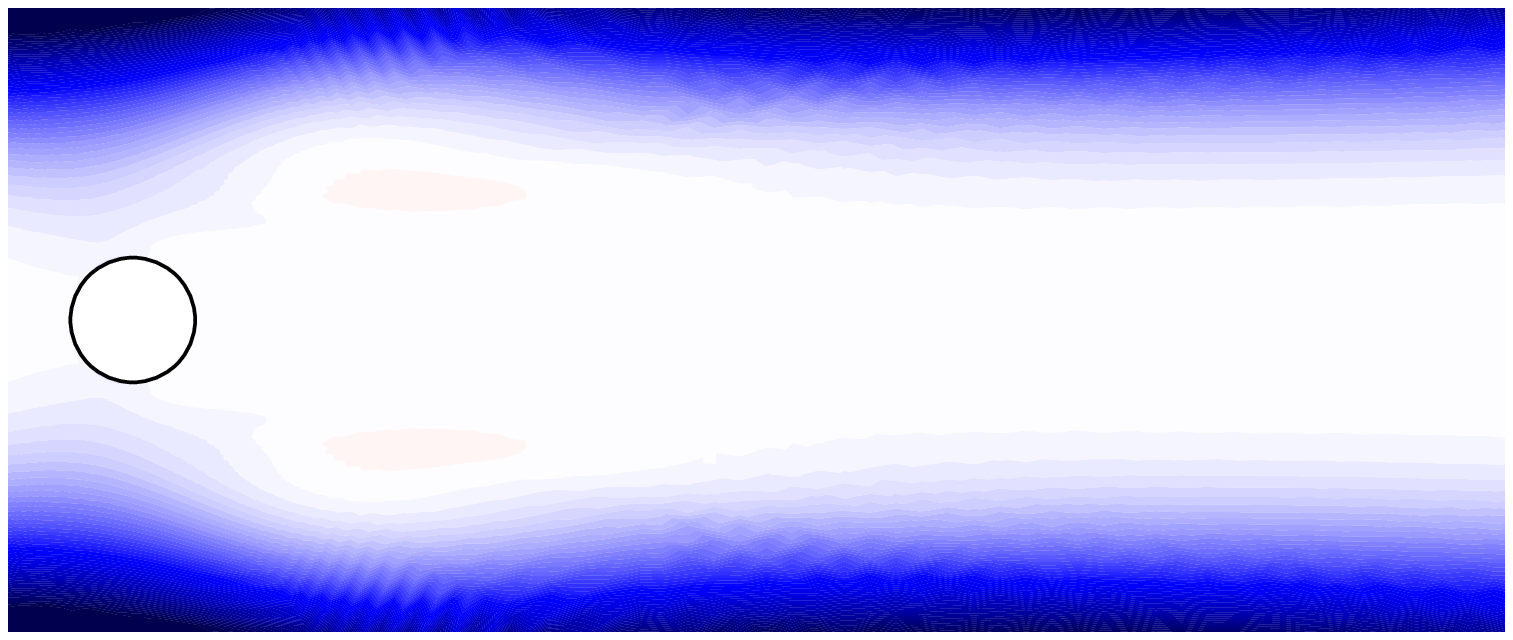} 
		& \includegraphics[width=0.2725\textwidth]{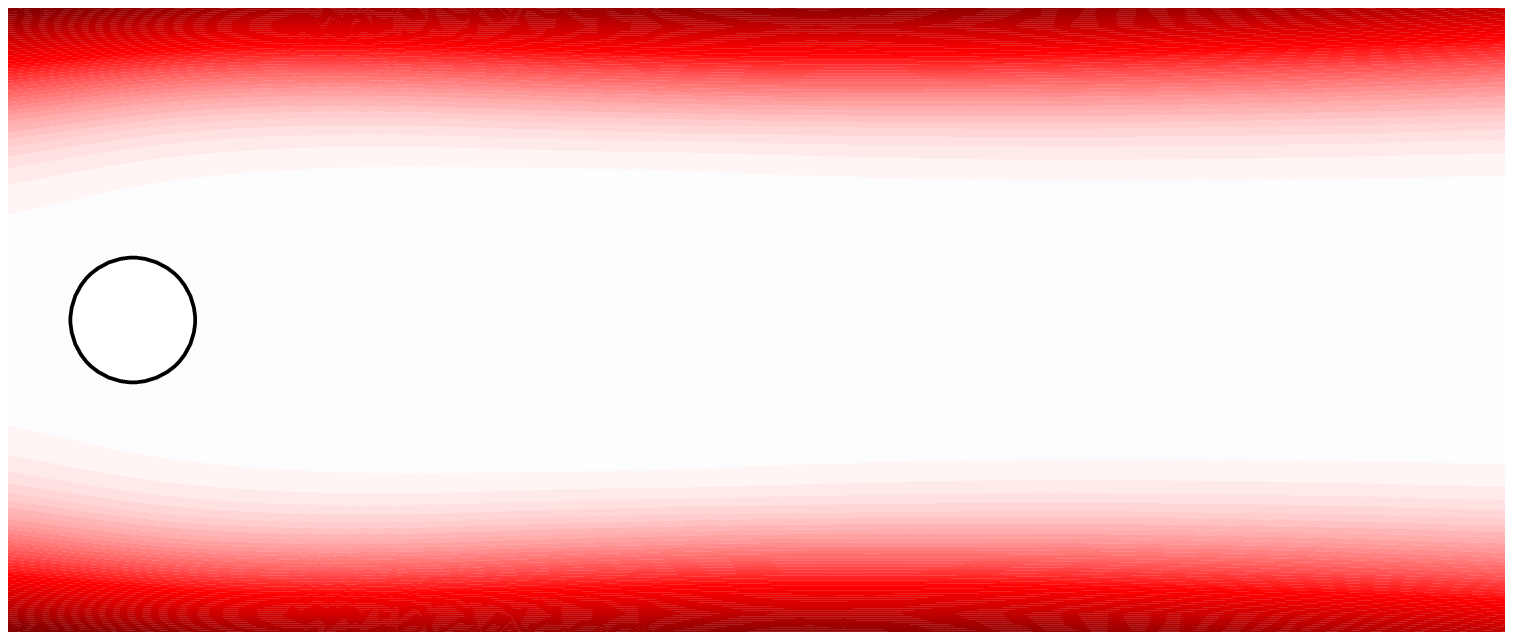}
		&\includegraphics[width=0.05\textwidth]{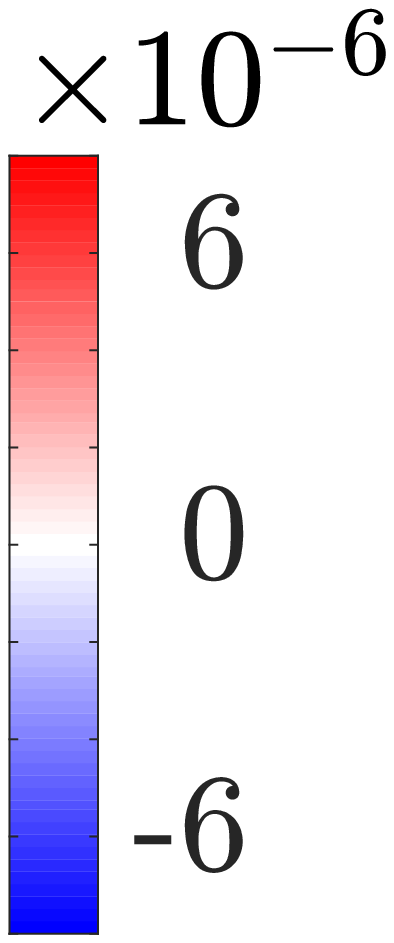}\\[2.5pt]
		\midrule
		&&&&\\[-0.8em]
		\llap{\parbox[b]{0in}{$\sum$\\\rule{0ex}{0.2in}}}& \includegraphics[width=0.2725\textwidth]{fig5a.eps} & \includegraphics[width=0.2725\textwidth]{fig5b.eps} 
		& \includegraphics[width=0.2725\textwidth]{fig5c.eps}
		&\includegraphics[width=0.05\textwidth]{fig5d.eps}\\[2.5pt]
		\midrule
		%\bottomrule 
	\end{tabular}
	\caption{\textcolor{black}{Spatial distribution of production $\hat{P}(\omega)$, linear energy dissipation $\hat{D}_e(\omega)$, and nonlinear energy transfer $\hat{N}(\omega)$ from resolvent analysis for each harmonic frequency. Note the smaller color scales for $\omega_2$ and $\omega_3$.}} \label{tab:svdcomponents}
\end{figure} 
\newcolumntype{C}{>{\centering\arraybackslash}m{0.45\textwidth}}
\aboverulesep=0ex
\belowrulesep=0ex
\begin{figure}
    \begin{center}
    	\begin{tabular}{m{7mm} | C | C }
		%\toprule
		$\ $ & Optimal Forcing Mode $\hat{\textbf{\textit{f}}}\hspace{0mm}^{(n)}_1$ & Optimal Response Mode $\hat{\textbf{\textit{u}}}\hspace{0mm}^{(n)}_1$\\[2.5pt]
		\midrule
		&&\\[-0.8em]
		\llap{\parbox[b]{0in}{$\omega_1$\\\rule{0ex}{0.2in}}}& \includegraphics[width=0.45\textwidth]{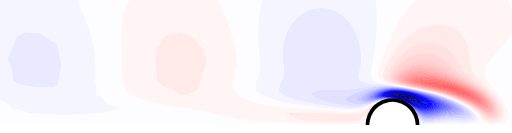} & \includegraphics[width=0.45\textwidth]{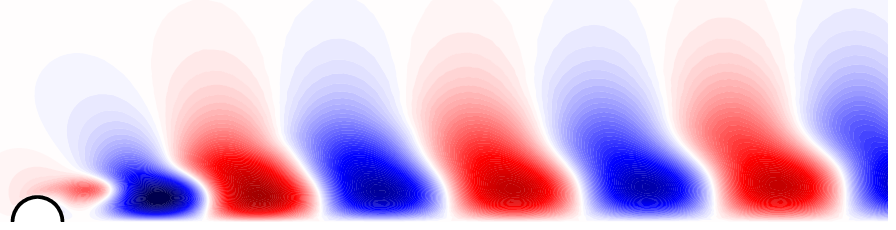}\\[2.5pt]
		\midrule
		&&\\[-0.8em]
		\llap{\parbox[b]{0in}{$\omega_2$\\\rule{0ex}{0.2in}}}& \includegraphics[width=0.45\textwidth]{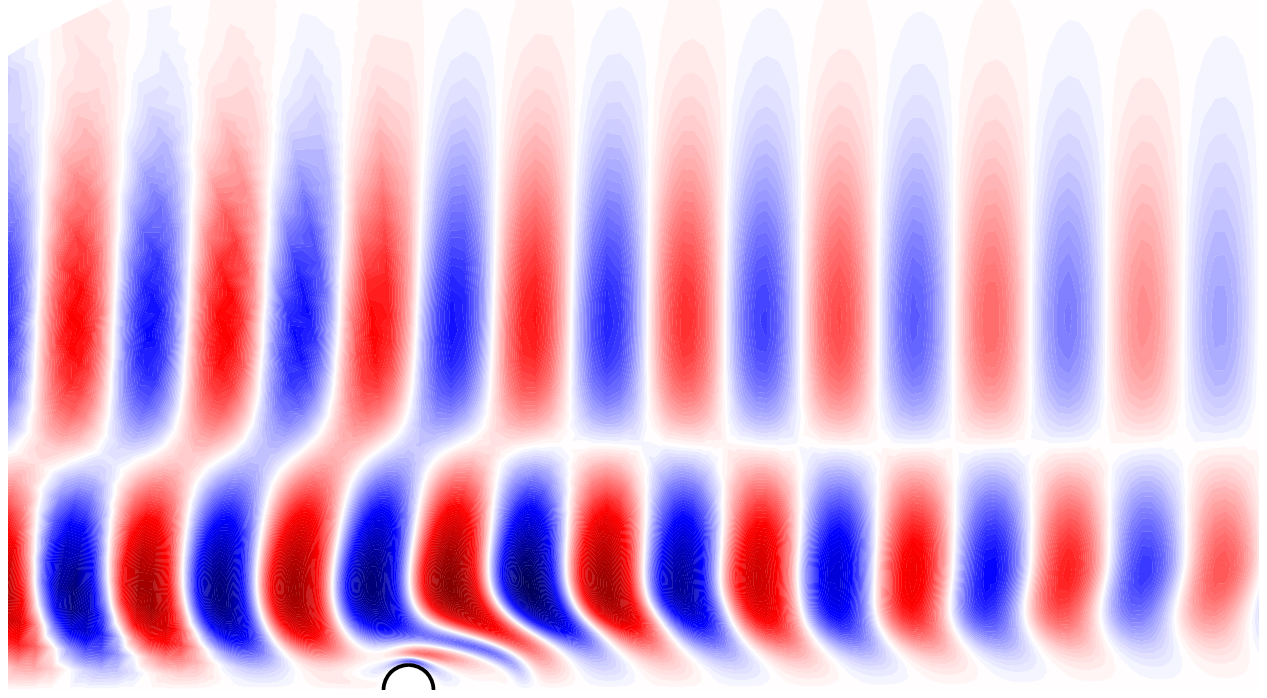} & \includegraphics[width=0.45\textwidth]{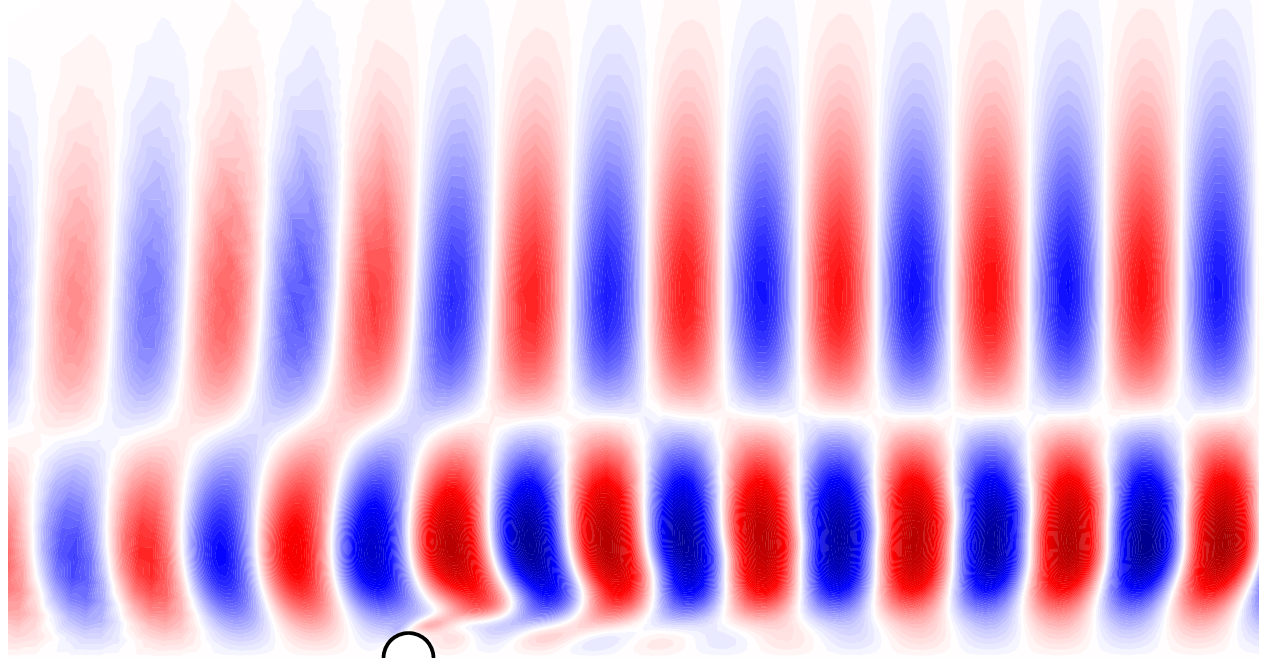}\\[2.5pt] 
		\midrule
		&&\\[-0.8em]
		\llap{\parbox[b]{0in}{$\omega_3$\\\rule{0ex}{0.2in}}}& \includegraphics[width=0.45\textwidth]{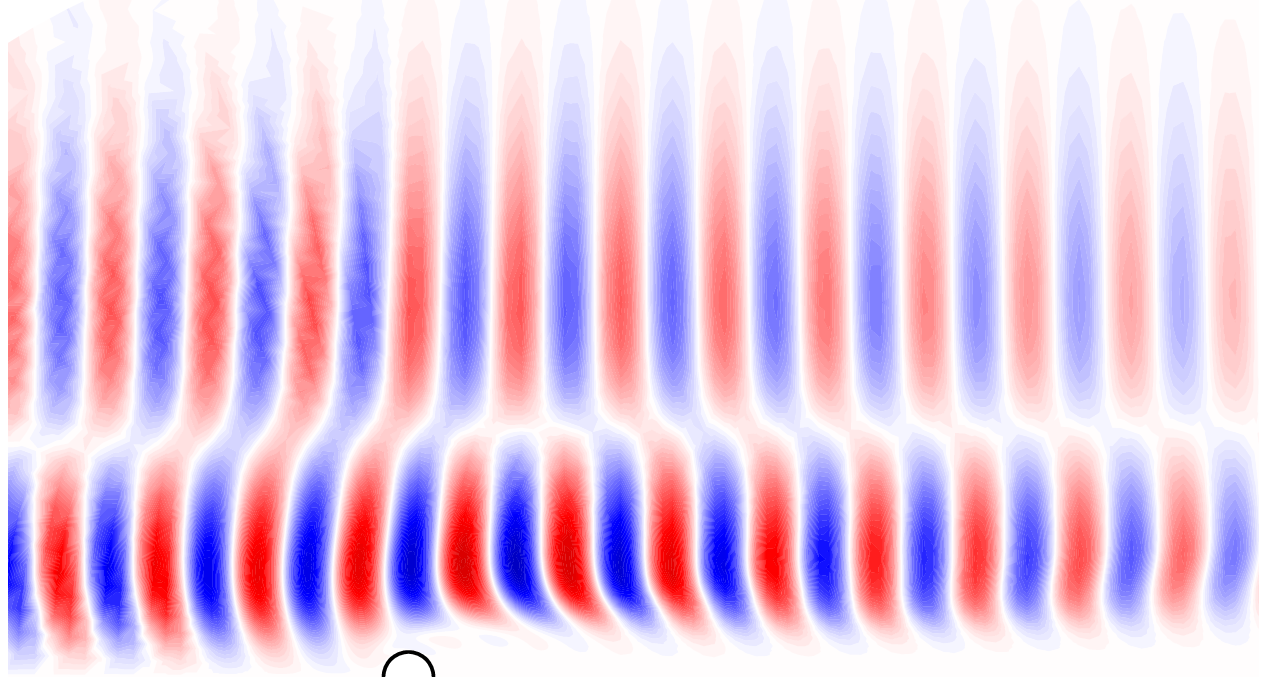} & \includegraphics[width=0.45\textwidth]{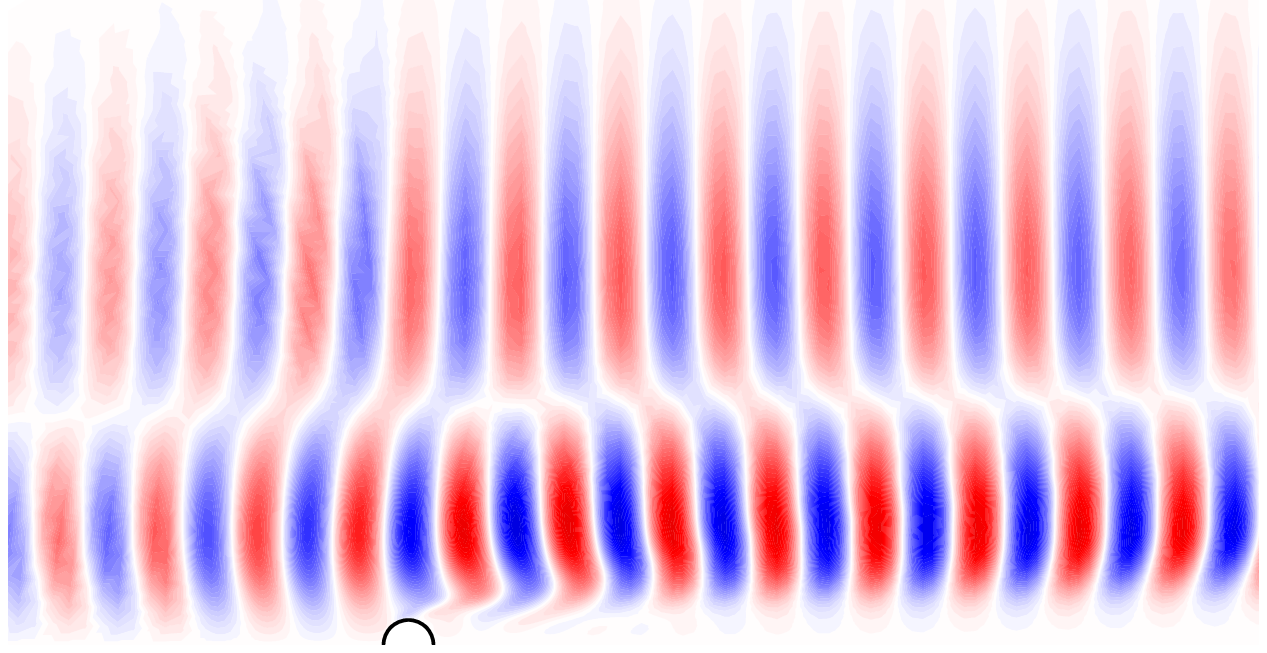}\\[2.5pt]
		\midrule
		%\bottomrule 
	\end{tabular}
    \end{center}
	\caption{\textcolor{black}{Optimal resolvent forcing modes and optimal resolvent response modes computed for $\omega_1=1.054$, $\omega_2=2.109$ and $\omega_3=3.163$ (real part, streamwise component, antisymmetric with respect to the centerline). A large upper half domain ($-8<x/D<17$, $0<y/D<15$) is shown for the higher harmonic modes to include the full spatial support of the structure.}} \label{fig:resolven_vel_highfre}
\end{figure}

Let us now consider the spatial distributions for resolvent analysis, which are plotted in Fig.~\ref{tab:svdcomponents}.
We see reasonable agreement between resolvent analysis and DNS for production and dissipation for the first harmonic mode and for the flow in aggregate.
The agreement for all other terms, however, is poor.
Particularly striking is that the nonlinear transfer term for the first harmonic, $\hat{N}(\omega_1)$, is approximately zero at all points in physical space.
Therefore not only is $\hat{N}(\omega_1)$ approximately zero when integrated in space (Figs.~\ref{fig:energybalance_comparison} \& \ref{fig:energybalance_dns}), but also at every point in space (Fig.~\ref{tab:svdcomponents}).
This implies that the inner product $\langle \hat{\textbf{\textit{f}}}\hspace{0mm}^{(1)}_1,\hat{\textbf{\textit{u}}}\hspace{0mm}^{(1)}_1 \rangle$ is approximately zero, which is consistent with the limited spatial overlap of $\hat{\textbf{\textit{u}}}\hspace{0mm}^{(1)}_1$ and $\hat{\textbf{\textit{f}}}\hspace{0mm}^{(1)}_1$ in physical space, \textcolor{black}{as shown by the first row in figure \ref{fig:resolven_vel_highfre},} and is explained by the the non-normality of the resolvent operator \citep{Chomaz05,symon2020energy}.
For the higher harmonics, we observe negligible energy production, dissipation, and nonlinear transfer in the wake. Instead, the spatial distribution of all three terms is concentrated in the free-stream. Moreover, it should be remarked that the colorbars for the resolvent distributions in rows 2 and 3 of Fig.~\ref{tab:svdcomponents} saturate at values that are two orders of magnitude smaller than those of the DNS in Fig.~\ref{tab:dnscomponents}.

\textcolor{black}{For a better physical understanding of the structures, we also present the optimal resolvent forcing and response modes in Fig.~\ref{fig:resolven_vel_highfre} for the harmonic frequencies $\omega_1$, $\omega_2$ and $\omega_3$. Each forcing-response pair achieves the maximum possible energy amplification at its corresponding harmonic frequency, as defined by the optimization problem \eqref{equ:resolvent_opti}. For higher harmonics (i.e.~$\omega_2$ and $\omega_3$),} it should be noted that the streamwise velocity has been plotted in a larger domain to include the free-stream part of the flow. Their optimal response fields consist of waves that are concentrated in the free-stream area of the flow. These flow structures are responsible for the free-stream advection of perturbations and their wavelength is inversely proportional to the oscillation frequency. \textcolor{black}{The spatial overlap of $\hat{\textbf{\textit{f}}}\hspace{0mm}^{(n)}_1$ and $\hat{\textbf{\textit{u}}}\hspace{0mm}^{(n)}_1$ in physical space is larger for the second and third harmonics. However, we see that this overlap is still small in the wake area since the strongest amplitudes for the response are downstream of the free-stream area and the strongest amplitudes for the forcing are upstream of the free-stream area.} Similar structures have been documented by \cite{dergham2013stochastic} for the flow over a backward-facing step. Consequently, we can identify two types of resolvent modes: i) those accounting for vortex shedding instabilities in the cylinder wake (e.g.~at $\omega_1$ shown in Fig.~\ref{fig:dnsresolvent}(d)) and ii) those representing the advection and diffusion of perturbations in the free-stream (e.g.~at $\omega_2$, $\omega_3$ shown in Fig.~\ref{fig:resolven_vel_highfre}). 
As we will show in the next section, the resolvent only models energy exchange between fluctuations and the mean flow. It is necessary to account for energy exchange between fluctuations and other fluctuations to correctly capture higher harmonics.

\section{Nonlinear energy transfer}\label{sec:result2}
\subsection{Triadic interactions and energy transfer}
The nonlinear term $N$, as well as the mean convection term $M$ in the fluctuation field, represent the mechanism by which fluctuation energy leaves the domain through the outlet boundary. However, when considered in the frequency domain, the nonlinear term $\hat{N}(\omega)$ is not a simple energy transportation mechanism for the flow as a whole but also an energy transfer mechanism among frequencies. For saturated vortex shedding, the fluctuation field can be considered as a summation of harmonic waves at several frequencies $\omega_n$, where the fluctuation energy is distributed as shown in  Fig.~\ref{fig:dnsresolvent}($\text{a}$). Substituting Eq.~\eqref{equ:fouriersummation} into the nonlinear forcing $\textbf{\textit{f}}'$ that stimulates fluctuations, we can expand the forcing $\hat{\textbf{\textit{f}}}$ as a series of nonlinear triadic interactions over several harmonic frequencies $\omega_n$:
\begin{equation}\label{equ:nonlinearscomps}
    \hat{f}_i^{(n)}=-\sum_{m\in\mathbb{Z}^{+}}(\hat{u}^{(n\textit{-}m)}_j\dfrac{\partial \hat{u}^{(m)}_i}{\partial x_j\ \ \ }+\hat{u}^{(n+m)}_j\dfrac{\partial \hat{u}^{(\textit{-}m)}_i}{\partial x_j\ \ \ }) \text{ with $m\neq n$},
\end{equation}
where $m,\ n\in\mathbb{Z}^{+}$ are positive integers. The superscript $(\cdot)^{(m)}$ represents the quantity at the $m$th harmonic frequency $\omega_m$ and $(\cdot)^{(\textit{-}m)}$ is its corresponding complex conjugate. It should be noted that setting $n = 0$ corresponds to the Reynolds stresses needed to sustain the mean flow, which is the zeroth harmonic (see Sec.~\ref{sec:n=0}). Furthermore, to avoid redundant formulations of the mean advection term $\textbf{\textit{U}}\cdot\nabla()$, $m$ can be any positive integer except $n$. 

For each integer $m$, we identify a pair of nonlinear interactions that generate an harmonic wave at frequency $\omega_n$: i) between two harmonic frequencies $\omega_{n-m}$ and $\omega_m$; ii) between two harmonic frequencies $\omega_{n+m}$ and $-\omega_m$. The work done by a general nonlinear forcing at $\omega_n$ can thus be written as:
\begin{gather}\label{equ:energytransfer}
    \begin{aligned}
        \underbrace{-\int (\hat{u}^{(n\textit{-}m)}_j\dfrac{\partial \hat{u}^{(m)}_i}{\partial x_j})\hat{u}_i^{(\textit{-}n)}\ d\Omega}_{\text{energy production at $\omega_n$}}&=\underbrace{\int (\hat{u}^{(n\textit{-}m)}_j\dfrac{\partial \hat{u}^{(\textit{-}n)}_i}{\partial x_j})\hat{u}_i^{(m)}\ d\Omega}_{\text{energy extracted from $-\omega_{m}$}}\underbrace{-\int (\hat{u}_i^{(m)}\hat{u}_i^{(\textit{-}n)}\hat{u}_j^{(n\textit{-}m)})\cdot\textbf{n} \ d\Gamma_{out}}_{\text{fluctuation energy flux out}},\\
        \underbrace{-\int (\hat{u}^{(n+m)}_j\dfrac{\partial \hat{u}^{(\textit{-}m)}_i}{\partial x_j})\hat{u}_i^{(\textit{-}n)}\ d\Omega}_{\text{energy production at $\omega_n$}}&=\underbrace{\int (\hat{u}^{(n+m)}_j\dfrac{\partial \hat{u}^{(\textit{-}n)}_i}{\partial x_j})\hat{u}_i^{(\textit{-}m)}\ d\Omega}_{\text{energy extracted from $\omega_{m}$}}\underbrace{-\int (\hat{u}_i^{(\textit{-}m)}\hat{u}_i^{(\textit{-}n)}\hat{u}_j^{(n+m)})\cdot\textbf{n} \ d\Gamma_{out}}_{\text{fluctuation energy flux out}},
    \end{aligned}
\end{gather}
where we have used the divergence theorem to convert integration over the domain to integration over the boundary. Now we can see that the work done by this pair of general nonlinear interactions at $\omega_n$ is equivalent to the negative work done by the corresponding interactions at $\pm\omega_m$ with the additional fluctuation energy leaving the domain. The balances reveal the energy transfer by triadic interactions among the recipient frequency $\omega_n$, the source frequency $\pm\omega_m$ and the intermediary frequency $\omega_{n\pm m}$ \citep{smyth1992spectral}. In other words, the harmonic wave at $\omega_n$ receives the energy extracted from harmonic frequency $\pm\omega_m$ with the help of intermediate advection at $\omega_{n\pm m}$. If the outlet boundary is infinitely far away from the cylinder, all fluctuations will eventually vanish:
\begin{equation}\label{equ:zerofluxout}
    -\int (\hat{u}_i^{(m)}\hat{u}_i^{(\textit{-}n)}\hat{u}_j^{(n\textit{-}m)})\cdot\textbf{n} \ d\Gamma_{out}=-\int (\hat{u}_i^{(\textit{-}m)}\hat{u}_i^{(\textit{-}n)}\hat{u}_j^{(n+m)})\cdot\textbf{n} \ d\Gamma_{out}=0\ \text{for $\Gamma_{out}\rightarrow \infty$},
\end{equation}
This is approximately true for the vortex shedding considered in this study, where the outlet boundary is located $23D$ downstream of the cylinder and the fluctuation energy flux out due to the nonlinearity is negligible (relative error $<0.5\%$). Thus, Eq.~\eqref{equ:energytransfer} describes an energy transfer mechanism that exchanges energy among frequencies. Importantly, these mechanisms are not revealed if we consider only a global energy balance as performed in Sec.\,\ref{sec:3.1}. 

\subsection{Nonlinear energy transfer among harmonic modes}
We consider the expansions of $\hat{N}(\omega_n)$, the work done by the nonlinearity at each harmonic frequency, as summations over different forcing components of Eq.~($\ref{equ:nonlinearscomps}$):
\begin{equation}
    \hat{N}(\omega_n)=\int (\hat{u}_i^{(\textit{-}n)}\hat{f}_i^{(n)}+\text{c.c.})\ d\Omega=\sum_{m\in\mathbb{Z}^{+}}\hat{\mathcal{N}}_{mn} \text{ with $m\neq n$},
\end{equation}
where $\hat{\mathcal{N}}_{mn}$ denotes the energy transfer mechanism from the $m$th harmonic frequency to the $n$th harmonic frequency. By implementing the assumption of zero fluctuation energy leaving the domain, as discussed in Eq.~\eqref{equ:zerofluxout}, the energy transfer mechanism $\hat{\mathcal{N}}_{mn}$ satisfies the relation:
\begin{gather}\label{eq:nonlintransfer_mn}
    \begin{aligned}
    \hat{\mathcal{N}}_{mn}=&-\int 2\Real\left\lbrace(\hat{u}^{(n\textit{-}m)}_j\dfrac{\partial \hat{u}^{(m)}_i}{\partial x_j})\hat{u}_i^{(\textit{-}n)}+(\hat{u}^{(n+m)}_j\dfrac{\partial \hat{u}^{(\textit{-}m)}_i}{\partial x_j})\hat{u}_i^{(\textit{-}n)}\right\rbrace\ d\Omega\\
    =&\int 2\Real\left\lbrace(\hat{u}^{(n\textit{-}m)}_j\dfrac{\partial \hat{u}^{(\textit{-}n)}_i}{\partial x_j})\hat{u}_i^{(m)}+(\hat{u}^{(n+m)}_j\dfrac{\partial \hat{u}^{(\textit{-}n)}_i}{\partial x_j})\hat{u}_i^{(\textit{-}m)}\right\rbrace\ d\Omega=-\hat{\mathcal{N}}_{nm}.
    \end{aligned}
\end{gather}
For the saturated vortex shedding, almost all the energy ($>99.9\%$) is concentrated in the first three harmonic frequencies. Thus, it is reasonable to truncate these expansions to the first three harmonics frequencies ($n,m\in[1,2,3]$). We may consider the nonlinear term $\hat{N}(\omega_n)$, therefore, to contain only a limited number of energy transfer terms $\hat{\mathcal{N}}_{mn}$:
\begin{subequations}\label{equ:nonlinearity}
    \begin{align}
        &\hat{N}(\omega_1)=\hat{\mathcal{N}}_{31}-\hat{\mathcal{N}}_{12},\\
        &\hat{N}(\omega_2)=\hat{\mathcal{N}}_{12}-\hat{\mathcal{N}}_{23},\\
        &\hat{N}(\omega_3)=\hat{\mathcal{N}}_{23}-\hat{\mathcal{N}}_{31},\\
        &N=\sum_{n=1}^{3}\hat{N}(\omega_n)=0.
    \end{align}
\end{subequations}
It should be noted that the energy balance across different frequencies, i.e.~Eq.~(\ref{equ:nonlinearity}\textit{d}), is consistent with the results in Fig.~\ref{fig:energybalance_dns}($\text{a}$). The detailed expressions for each energy transfer term among the first three harmonic frequencies can be formulated as:
\begin{subequations} \label{eq:exchange}
    \begin{align}
        \nonumber
        \hat{\mathcal{N}}_{12}=&-\int2\Real\left\lbrace
        \begin{bmatrix}
        \hat{u}^{(1)}_j&\hat{u}^{(\textit{-}3)}_j
        \end{bmatrix}
        \dfrac{\partial \hat{u}^{(1)}_i}{\partial x_j}
        \begin{bmatrix}
        \hat{u}_i^{(\textit{-}2)}\\
        \hat{u}_i^{(2)}
        \end{bmatrix}\right\rbrace\ d\Omega\\
        =&\int2\Real\left\lbrace
        \begin{bmatrix}
        \hat{u}^{(\textit{-}1)}_j&\hat{u}^{(\textit{-}3)}_j
        \end{bmatrix}
        \dfrac{\partial \hat{u}^{(2)}_i}{\partial x_j}
        \begin{bmatrix}
        \hat{u}_i^{(\textit{-}1)}\\
        \hat{u}_i^{(1)}
        \end{bmatrix}\right\rbrace\ d\Omega,\\
        \hat{\mathcal{N}}_{23}=&-\int2\Real\left\lbrace(\hat{u}^{(1)}_j\dfrac{\partial \hat{u}^{(2)}_i}{\partial x_j})\hat{u}_i^{(\textit{-}3)}\right\rbrace\ d\Omega=\int2\Real\left\lbrace(\hat{u}^{(\textit{-}1)}_j\dfrac{\partial \hat{u}^{(3)}_i}{\partial x_j})\hat{u}_i^{(\textit{-}2)}\right\rbrace \ d\Omega,\\
        \hat{\mathcal{N}}_{31}=&-\int2\Real\left\lbrace(\hat{u}^{(\textit{-}2)}_j\dfrac{\partial \hat{u}^{(3)}_i}{\partial x_j})\hat{u}_i^{(\textit{-}1)}\right\rbrace \ d\Omega=\int2\Real\left\lbrace(\hat{u}^{(2)}_j\dfrac{\partial \hat{u}^{(1)}_i}{\partial x_j})\hat{u}_i^{(\textit{-}3)}\right\rbrace \ d\Omega,
    \end{align}
\end{subequations}
where we have classified three pairs of triadic interactions that only involve the first three harmonic frequencies, as derived in Eq.~\eqref{equ:energytransfer} under the assumption that the flux out is negligible. Note that $\hat{\mathcal{N}}_{23}$ and $\hat{\mathcal{N}}_{31}$ have only one term due to the truncation of the expression so that the intermediary frequency $m+n\leq 3$. If more harmonics had been retained, they would contain two terms like $\hat{\mathcal{N}}_{12}$. 

\begin{figure}
	\centerline{\includegraphics[width=0.95\textwidth]{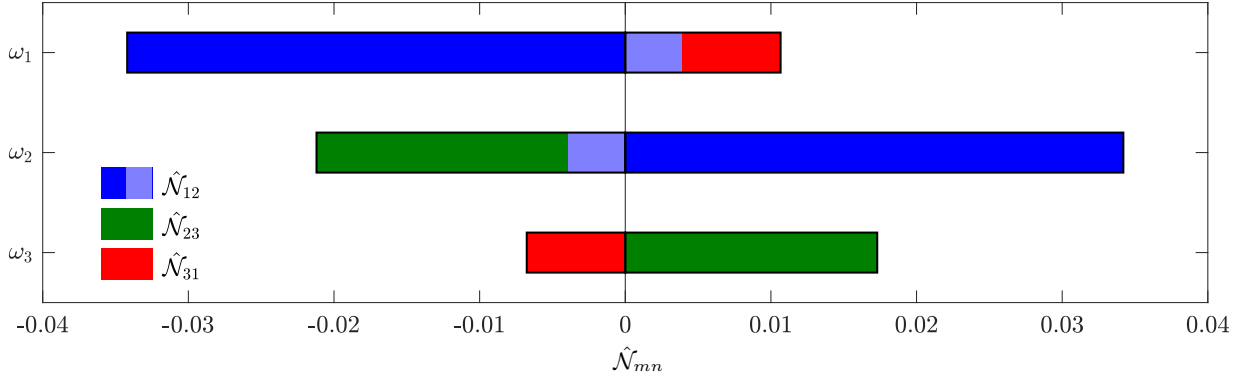}}
	\centerline{\llap{\parbox[b]{0in}{$\hat{\mathcal{N}}_{mn}$\\\rule{0ex}{0in}}}}
	\vspace{-5mm}
	\caption[Nonlinear energy transfer over the first three harmonic frequencies]{Nonlinear energy transfer mechanisms $\hat{\mathcal{N}}_{mn}$ over the first three harmonic frequencies.}
	\label{fig:energytransfer_dns}
\end{figure}

The nonlinear energy exchanges from Eq.~\ref{eq:exchange} are illustrated in Fig.~\ref{fig:energytransfer_dns}. Unlike Fig.~\ref{fig:energybalance_comparison}, which illustrates the net energy gain or loss due to nonlinear transfer, Fig.~\ref{fig:energytransfer_dns} shows that multiple nonlinear exchanges exist for each harmonic frequency. We first consider $\hat{\mathcal{N}}_{12}$, which consists of two terms and denotes transfer between $\omega_1$ and $\omega_2$. The first term represents the energy cascade from low to high frequencies. Energy is removed from $\omega_1$, advected by $\omega_1$, and transferred to $\omega_2$ as denoted by the dark blue bar in Fig.~\ref{fig:energytransfer_dns}. There also exists feedback from $\omega_2$, which returns energy to $\omega_1$ as indicated by the light blue bar in Fig.~\ref{fig:energytransfer_dns}. For this exchange $\omega_3$ is the intermediary frequency. The second term $\hat{\mathcal{N}}_{23}$ consists of one term where energy is transferred from $\omega_2$ to $\omega_3$ (green bar). Finally, $\hat{\mathcal{N}}_{31}$ consists of one term where energy is transferred from $\omega_3$ to $\omega_1$ (red bar). Interestingly, $\omega_1$ does not directly give energy to $\omega_3$ since the number of harmonics considered is limited to $n+m \leq 3$. Instead $\omega_3$ receives its energy from $\omega_2$.

\subsection{Nonlinear energy transfer from the mean flow} \label{sec:n=0}
We now consider a special case where the triad interactions transfer energy from the mean flow, or the zeroth harmonic, to harmonic waves. Note that if $m=0$, Eq.~\eqref{eq:nonlintransfer_mn} reduces to the energy production term $\hat{P}(\omega_n)$:
\begin{gather}\label{eq:nonlintransfer_mean}
    \begin{aligned}
    \hat{P}(\omega_n)=\hat{\mathcal{N}}_{0n}=&-\int 2\Real\left\lbrace(\hat{u}^{(n)}_j\dfrac{\partial U_i}{\partial x_j})\hat{u}_i^{(\textit{-}n)}\right\rbrace\ d\Omega\\
    =&\int 2\Real\left\lbrace(\hat{u}^{(n)}_j\dfrac{\partial \hat{u}^{(\textit{-}n)}_i}{\partial x_j})U_i\right\rbrace\ d\Omega=-\hat{\mathcal{N}}_{n0},
    \end{aligned}
\end{gather}
where $\hat{\mathcal{N}}_{n0}$, which models the work done by the partial Reynolds stress forcing resulting from the $n$th harmonic wave, is equivalent to the energy produced at the harmonic frequency $\omega_n$. As discussed in Sec.~\ref{sec:balDNS}, energy production $\hat{P}(\omega_n)$ is positive as energy is removed from the mean flow and given to the harmonic waves without any feedback. We recall that most of the energy from the mean flow is transferred to the first harmonic frequency \textcolor{black}{($\hat{P}(\omega_1)\neq0$)}, as presented in Fig.~\ref{fig:energybalance_comparison}(\textit{a}). As a result, the corresponding energy production at higher harmonic frequencies is almost negligible ($\hat{P}(\omega_2)\approx 0$, $\hat{P}(\omega_3)\approx 0$). 

\textcolor{black}{The principal source of energy for higher harmonic modes is therefore the fundamental harmonic mode rather than the mean flow. Moreover, the corresponding energy production terms for the higher harmonics $\hat{P}(\omega_2)$ and $\hat{P}(\omega_3)$ are small in Fig.~\ref{fig:energybalance_comparison}(\textit{a}), signifying that they extract little energy from the mean flow.} This has two implications. First, the fundamental harmonic wave is responsible for the saturation of the mean flow, which explains the success of self-consistent models that consider interactions between the mean flow and the fundamental harmonic only \cite{mantivc2014self, mantivc2015self,Turton15}. Second, resolvent analysis is not able to predict the structure of the higher harmonics since the principal source of their energy is not the mean flow. The resolvent is similar to a quasilinear (QL) approximation \cite{Farrell07, Marston08, Srinivasan12} in that it models coupling between the mean and fluctuations. It does not directly consider interactions between fluctuations and other fluctuations. \textcolor{black}{Therefore, resolvent analysis correctly predicts that the energy production terms for the higher harmonic modes are negligible. However, the resolvent-based prediction is not accurate since these higher harmonic modes arise not from linear amplification mechanisms but from a direct energy injection from the fundamental harmonic mode.}

\subsection{Energy cascade and modeling}
\begin{table}
	\begin{ruledtabular}
		\begin{tabular}{lccccc}
    		$\hat{\mathcal{N}}_{mn}$ & Advection & Source & Recipient & \textcolor{black}{Magnitude ($\times10^{-2}$)} &Cascade \\
			\hline
        	$\hat{\mathcal{N}}_{01}$ ($\hat{P}(\omega_1)$) & 1 & 0 & 1 & \textcolor{black}{24.2} &Down  \\
        	$\hat{\mathcal{N}}_{12}$ & 1 & 1 & 2 & \textcolor{black}{3.42} &Down \\
        	\textcolor{white}{$\hat{\mathcal{N}}_{12}$} & 3 & 2 & 1 & \textcolor{black}{0.39} &Up \\
        	$\hat{\mathcal{N}}_{23}$ & 1 & 2 & 3 & \textcolor{black}{1.73} &Down \\
        	$\hat{\mathcal{N}}_{31}$ & 2 & 3 & 1 & \textcolor{black}{0.68}&Up \\
		\end{tabular}
	\end{ruledtabular}
	\caption{\label{tab:triads} Energy transfer due to triadic interactions in cylinder flow.}
\end{table}

The relevant nonlinear energy exchanges taking place in the cylinder flow are summarized in Table~\ref{tab:triads}. Energy is removed from the source mode and transferred to the recipient mode. An advection mode is necessary to facilitate each transfer although it is passive, i.e.~it does not supply or receive energy. \textcolor{black}{The final two columns indicate the direction of the spectral transfer mechanisms, as discussed by \cite{smyth1992spectral}, for flow structures at different temporal frequencies.} If the harmonic frequency of the source mode is less than that of the recipient mode, energy is cascading down from large scales (or lower harmonic frequencies) to the small scales (higher harmonic frequencies). If, on the other hand, the harmonic frequency of the source mode is greater than that of the recipient mode, energy is being fed back (or cascaded up) from small scales (higher harmonic frequencies) to large scales (lower harmonic frequencies). \textcolor{black}{Although the contributions of any cross-frequency energy transfers are relatively small compared to the mean energy production (i.e.~$\hat{\mathcal{N}}_{01}$), they are critical for determining the spatial distribution of the harmonic modes. It is worth noting that, for this flow, the energy transfer from the small scales to the large scales (rows 3 and 5 of Table~\ref{tab:triads}) is not negligible, especially for the third harmonic which transfers around $40\%$ of the total energy it receives (from the second harmonic) back to the fundamental mode.}

\textcolor{black}{
The prediction of the harmonic mode shapes can be improved by either directly solving a subset of nonlinear interactions (i.e.~Eq.~\eqref{equ:nonlinearity}), using the harmonic balance method \citep{rigas2020non}, or by modeling the nonlinear forcing rather than considering it as an unknown forcing. In particular, a recent study concerning the cylinder wake formulated an extended operator by including an additional model using the knowledge of the second-harmonic that contains an eigenmode matching the structure of the fundamental vortex shedding mode \citep{marquet2020extended}. Similarly, an appropriate eddy viscosity model may lead to an improvement in the structural prediction of the fundamental mode (or the most energetic mode) \citep{Mettot14}.  However, eddy viscosity is less applicable for higher harmonics that receive energy through nonlinear transfer since it introduces additional dissipation that can only remove energy from the flow \citep{symon2020energy}. This limitation can be overcome by formulating a harmonic resolvent operator about a time-varying base flow, which models nonlinear energy transfer mechanisms and reveals important information concerning disturbance amplification across scales and frequencies \citep{padovan_otto_rowley_2020}. In this case, linear modeling of nonlinear energy transfer requires knowledge of not only the time-averaged mean flow but also the dominant harmonic components. Indeed, forming a linear operator that includes cross-frequency mechanisms requires additional information rather than only the time-invariant base flow. That is, the modeling of nonlinear interactions for the large scales (lower harmonic frequencies) exploits the spatial structure of the small scales (higher harmonic frequencies) and the modeling for the small scales (higher harmonic frequencies) requires large-scale information (lower harmonic frequencies). The current work has clarified the contribution of each nonlinear interaction and revealed the corresponding direction of energy transfer. A full characterization of the nonlinear interactions may exploit the structure of the extended resolvent operator (e.g.~the mixed formulation that consists of several dominant harmonics) which provides a view to self-consistent, nonlinear, low-order models for fluid flows.
}

\section{Conclusions}\label{sec:conclusion}
We have investigated energy transfer mechanisms for vortex shedding behind a 2D circular cylinder at $\Rey = 100$. An energy balance is achieved across production, viscous dissipation, and nonlinear transfer both for the flow as a whole and for each harmonic mode. Production is generally positive and extracts energy from the mean flow, whereas viscous dissipation is always negative. Meanwhile, nonlinear mechanisms transfer energy between temporal frequencies to ensure an energy balance for each harmonic mode. Specifically, the nonlinear energy transfer is negative for the fundamental harmonic frequency $\omega_1$ but positive for its higher harmonics. \textcolor{black}{The net energy transfer across all harmonics is approximately zero because the nonlinear terms are conservative, and there exists an energy balance across all linear mechanisms for the flow as a whole, which is consistent with the Reynolds-Orr equation for localized or spatially periodic fluctuations.}

The energy balance achieved by the DNS was compared to that predicted by resolvent analysis. The resolvent operator, when forced at each harmonic frequency by its leading forcing mode, achieves an energy balance for each harmonic mode, and for the flow in aggregate. Although reasonable agreement was observed between resolvent analysis and DNS for production and dissipation for the fundamental harmonic wave, it does not achieve a suitable balance for the nonlinear transfer of energy across harmonic waves. In particular the nonlinear transfer of energy from the first harmonic frequency to the second and third harmonic frequencies seen for the DNS is not captured by resolvent analysis. Instead, it identified linear mechanisms that account for the advection and diffusion of perturbations in the free-stream at higher harmonic frequencies. This helps to explain the excess energy observed in the fundamental harmonic frequency's leading resolvent mode (Fig.\,\ref{fig:dnsresolvent}\,(\textit{d})) when compared to the true fundamental harmonic mode shape (Fig.\,\ref{fig:dnsresolvent}\,(\textit{c})). 

Detailed energy exchange mechanisms were revealed by expanding the nonlinear terms $\hat{N}({\omega_n})$ into a series of triadic interactions. These appear in pairs and transfer energy from a source frequency to a recipient frequency with the help of an intermediary frequency \cite{smyth1992spectral}. The nonlinear terms account for not only energy cascades from low frequencies to high frequencies, but also a considerable amount of energy feedback from high frequencies to low frequencies. The analysis was also extended to the energy transfer between the mean flow,  or the zeroth harmonic wave, and the other harmonics. The fundamental harmonic wave is the principal recipient of energy from the mean flow unlike higher harmonic waves whose principal source of energy is the fundamental harmonic wave. This explains the failure of the standard resolvent analysis (based on the mean flow alone) on predicting higher harmonics waves. Together these observations provide necessary prerequisites and constraints for developing eddy-viscosity models to model nonlinear energy transfer and understand the limitations of quasilinear approximations of unsteady flows.
%\nocite{*}

\bibliography{bibliography}% Produces the bibliography via BibTeX.

\end{document}